# LiNiO2/NiO Phase Prediction Using Artificial Neural Networks

Jonas Scheunert[1,2], Shamail Ahmed[1,2], Thomas Demuth[1,2], Andreas Beyer[1,2] and Kerstin Volz[1,2*]

[1]Philipps-Universität Marburg, mar.quest | Marburg Center for Quantum Materials and Sustainable Technologies, Hans-Meerwein-Straße 6, 35032 Marburg, Germany

[2]Philipps-Universität Marburg, Department of Physics, Hans-Meerwein-Straße 6, 35032 Marburg, Germany

[*]Corresponding author: kerstin.volz@physik.uni-marburg.de

## Abstract

In the realm of material analysis, identifying different material phases is of key importance. Artificial intelligence in the form of neural networks provides a very fast and, once trained, computationally inexpensive method for analysing large amounts of image data, like the sets of diffraction patterns generated during four-dimensional scanning transmission electron microscopy (4DSTEM) dataset acquisition. In this work, we train multiple network architectures on images of this type to distinguish between the phases of LiNiO2 and NiO, an important and challenging distinction in the lithium-ion battery community, since the formation of NiO limits battery capacity. We test both classical convolutional neural networks (CNNs) and different forms of vision transformers (ViTs). Our networks are trained on synthetic images and tested on experimentally recorded diffraction patterns. Additionally, we also investigate the decision-making of our networks using the GradCAM method. We test our networks on both synthetic as well as experimental diffraction patterns. Our networks exhibit very robust results, especially when dealing with highly varying data, an area where traditional template-matching methods typically struggle.

## Introduction

Advancements in data acquisition methods for material analysis have fundamentally improved the quality and the amount of information that can be gained by investigating images. Among these methods, four-dimensional scanning transmission electron diffraction (4DSTEM) has emerged as a powerful tool for probing local crystal structures with high spatial resolution [1-4]. By recording a spot-based nano-beam diffraction pattern at every probe position during a scan, 4DSTEM datasets provide detailed information about crystallographic orientation and phase composition of a sample. However, the large amount of diffraction data generated by such experiments presents a significant challenge for conventional analysis methods, which often require extensive computational resources as well as substantial user interaction and experience.

Phase identification using electron diffraction patterns is a particularly important task in the characterization of functional materials. In battery materials, for example, the formation of secondary phases can strongly influence electrochemical performance, stability, and degradation behavior [5]. A notable example is the coexistence of $LiNiO_2$ (LNO) and rock-salt NiO phases. While $LiNiO_2$ is a technologically relevant cathode material for lithium-ion batteries, the formation of NiO is often associated with degradation processes and performance loss [6, 7]. Reliable differentiation between these two phases is therefore of considerable practical importance. However, due to their structural similarity, their diffraction patterns also exhibit strong similarities for many crystallographic orientations, making phase discrimination challenging even for established diffraction analysis approaches.

Traditional methods for diffraction pattern analysis commonly rely on template matching with large banks of simulated diffraction patterns [8]. While these techniques can achieve excellent accuracy under ideal conditions, they often require precise knowledge of experimental parameters such as camera length, beam energy, material thickness, convergence angles, etc. In addition, the user is typically required to manually align and significantly post-process the experimental patterns for the template-matching software to suppress noise and match them to simulations. However, small intensity variation, calibration errors, distortions, or variations in other imaging conditions can significantly affect the performance of a traditional template-matching process. Furthermore, the computational cost of matching large numbers of experimental diffraction patterns against extensive libraries of simulated templates can become prohibitive for large 4DSTEM datasets, even for kinematically simulated patterns.

Recent advances in machine learning, particularly deep neural networks, offer an attractive alternative. Once trained, neural networks can classify images with high speed and comparatively low computational cost, making them well-suited for the analysis of large diffraction datasets. Convolutional neural networks (CNNs) have demonstrated remarkable success in a wide range of image recognition tasks and have already been applied to several problems in electron microscopy [9][10]. More recently, transformer-based architectures, originally developed for natural language processing, have been adapted for image analysis [11]. Vision Transformers (ViTs) and their variants employ self-attention mechanisms to model long-range relationships within images and have achieved state-of-the-art performance in many computer vision benchmarks. Hybrid architectures combining CNNs and ViTs have also shown promise by exploiting both local feature extraction and global contextual information [12].

Despite these advances, the application of transformer-based architectures to diffraction pattern analysis remains largely unexplored. In particular, it is not yet clear how their performance compares with conventional CNNs for challenging phase discrimination tasks involving highly similar diffraction signatures. Moreover, the ability of such networks to generalise from simulated training data to experimentally acquired diffraction patterns remains an important open question.

In this work, we investigate the applicability of several deep learning architectures for phase identification in 4DSTEM diffraction data. We compare classical convolutional neural networks, ViTs, hybrid CNN–ViT models, and Swin-Transformers [13] on the task of distinguishing $LiNiO_2$ and NiO diffraction patterns. Additionally, we investigate the validity of using separate networks, trained to predict the orientation of either LNO or NiO phases, and comparing their confidences to determine a phase’s presence. All networks are trained exclusively on large-scale synthetic diffraction datasets generated from Bloch-wave simulations implemented in py4DSTEM [14] and are subsequently evaluated using both simulated test data and experimentally recorded diffraction patterns. To assess robustness, we investigate the influence of severe image distortions and previously unseen crystal orientations. Furthermore, we employ Gradient-weighted Class Activation Mapping (GradCAM) [15] to gain insight into the decision-making processes of the most successful network architectures and to identify the diffraction features that drive phase classification. Through this comparison, we aim to evaluate the suitability of modern deep learning approaches for automated phase mapping in electron diffraction experiments and to provide guidance for future applications of artificial intelligence in transmission electron microscopy.

## Results

We follow two separate approaches to use neural networks for crystal phase prediction. Firstly, two separate CNNs are trained, one on LNO and one on NiO data, to predict the orientations of diffraction patterns for the two materials separately. For this, every training diffraction pattern is given a label consisting of the three Euler angles in the Bunge notation ([phi1, PHI, phi2]). The phase of a diffraction pattern of an unknown sample is then later determined by comparing the confidences of the two separate networks when predicting its orientation. The second approach followed are dedicated phase networks, which are trained on patterns of both materials. Here the training patterns are labeled with the material phase (LNO, NiO, amorphous or vacuum). The architectures tested are a classic CNN, a ViT as well as two hybrid networks that combine aspects of both network types. Lastly a Swin-Transformer was also analysed.

### Predictions on simulated data

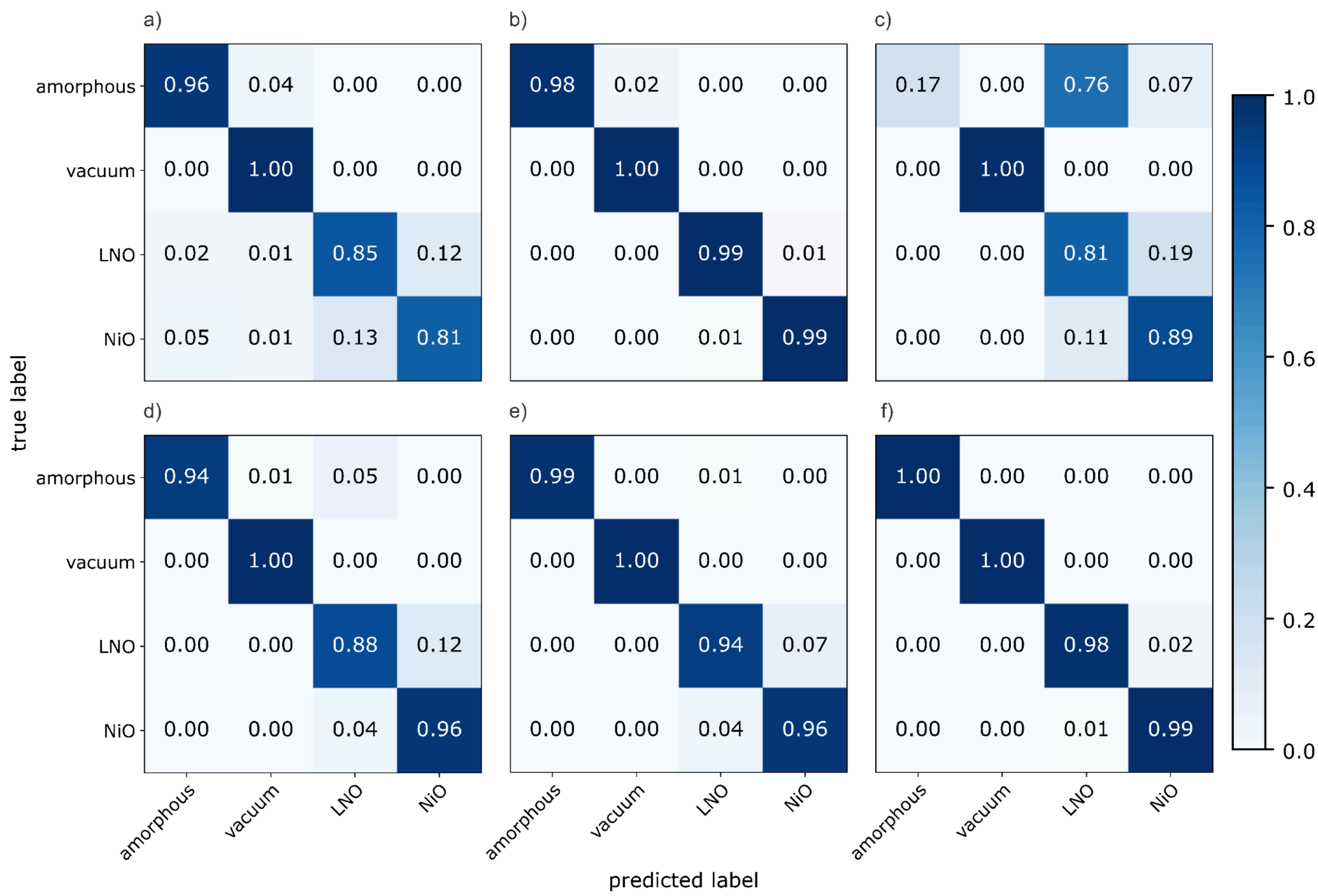

Figure 1: **Confusion matrices for the networks created by classifying diffraction patterns of the training dataset.** For each phase (LNO, NiO, amorphous and vacuum), 1000 randomly selected images are predicted. a) Phase predictions achieved by comparing the rescaled confidences of two orientation networks that predict LNO and NiO, respectively. b) - f) Phase prediction by dedicated phase networks (CNN, ViT, two CNN-ViT hybrids and Swin-T). The second CNN-ViT hybrid model combines the input image with the final CNN feature map to create the input for the transformer block.

To get an initial impression of how well our networks learned to differentiate the phases, we create confusion matrices using the original training images. Of each phase (LNO, NiO, vacuum, and amorphous), 1000 randomly selected images are given to the networks. Figure 1 shows the confusion matrices for both the phase-prediction networks as well as for the case of two networks predicting the orientations of NiO and LNO, whose confidences are compared to each other in order to determine the phase from the diffraction pattern. It becomes obvious that the classic convolutional neural network, as well as the Swin-Transformer model, are the most successful in this regard. These networks are able to differentiate all phases of the training data with more than 97% accuracy. The vision transformer model is clearly less capable in this regard. It is only able to classify around 80% of the LNO and NiO images correctly. As would be expected, the incorrectly predicted images are all categorised to belong to the other material system instead of the vacuum or amorphous class. However, the vision transformer also struggles with images of the amorphous category. Interestingly, most of these are predicted to belong to the LNO class. Combining the vision transformer with a small CNN block reduces both shortcomings of this network. Further changing the input of the transformer part of the network, to receive both the feature maps of the CNN block as well as the original diffraction pattern, improves the results to almost rival our other networks. This network architecture is able to achieve an accuracy of 94%. Finally, the method of comparing the outputs of two separate orientation CNNs results in weaker but still promising predictions. While both the vacuum and amorphous phases are categorised almost perfectly, about 12% of images for both the NiO and LNO phases are predicted to be the wrong material phase with this method.

One of the biggest challenges to the usage of conventional template-matching algorithms for any kind of analysis is the necessity of correct user inputs. Those programs need very accurate information about the experimental parameters of the microscope, like the camera length and the diffraction pattern's geometric distortions. In addition, it is necessary to precisely align the diffraction pattern with the internal templates of the program. This alignment step is typically only done once for a whole dataset, although it may change over the field of view, e.g., due to an unavoidable Descan in 4DSTEM measurements. Accordingly, misalignments are commonplace. Figure 2 shows how the predictions of the reference program change when the alignment is slightly shifted (b) or when a slightly wrong camera length is used (c). For this, we let a conventional template-matching software (ASTAR) [16] predict the phase of an unaugmented synthetic LNO dataset. To analyse our networks' capabilities in dealing with such potential user errors, we let them predict challenging simulated diffraction patterns with cropping and shifting far beyond the values used in the training data. This simulates a potential Descan as well as a varying camera length. Supplementary Figure S1 shows examples of such diffraction patterns. The diffraction patterns used in this test encompass the entire set of orientations in the training dataset, excluding in-plane rotations by a varying phi1 angle. Consequently, one can plot the networks' predictions directly onto an Inverse Pole Figure (IPF) and get an assessment of potentially problematic areas of orientations for the networks. In Figure 3 such IPFs are shown for the CNN, the CNN-ViT hybrid and Swin-Transformer phase-networks for both tested types of augmentation of LNO patterns. Additional graphics of the predictions of these and other network architectures predicting patterns of LNO and NiO are shown in the Supplementary Figures S2 and S3. The results show that our networks are clearly more stable to a shifting diffraction pattern as opposed to a change in the camera length. This is in line with our results for comparable tests using template-matching software. The CNN-ViT hybrid model and Swin-Transformer show very similar results. For both architectures the diffraction patterns close to the [0001]-orientations are the most problematic. At very high levels of cropping, this challenging area quite suddenly starts to expand across the whole IPF. In contrast to those networks, the CNN shows a truly remarkable resilience to the chosen augmentations. It demonstrates perfect labeling for all orientations, regardless of any amount of cropping or shifting of the diffraction pattern. A possible explanation for the weaker results of the transformers may be their typical strength, namely, their ability to recognise global relationships, i.e., between very distant diffraction spots. For increasingly strong cropping, high-angle diffraction spots become no longer visible and therefore unavailable for networks that rely on them. In contrast, CNNs focus much more on local structures, which are largely unaffected by such cropping.

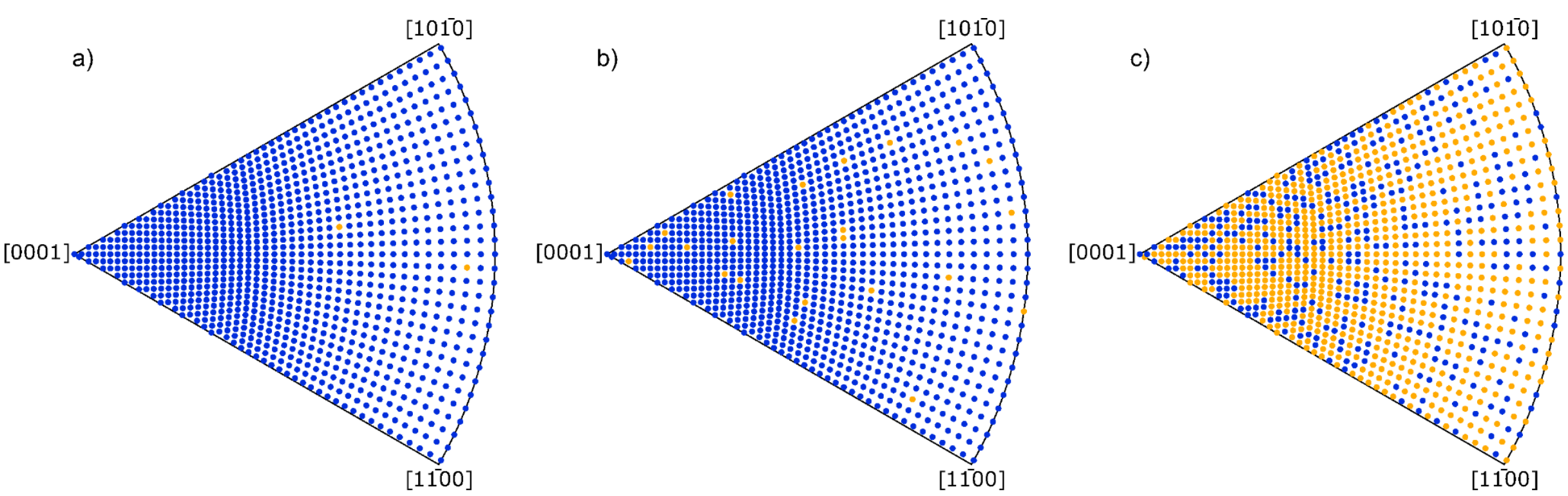

Fig 2: **Comparison of the predictions of a template-matching program for all simulated orientations (excluding variation in phi1) of LNO for correct and wrong alignments.** a): correct alignment, b) pattern center shifted by 5 pixels with respect to the simulations, c) wrong camera length calibration. Blue: correct LNO prediction, Orange: false NiO prediction

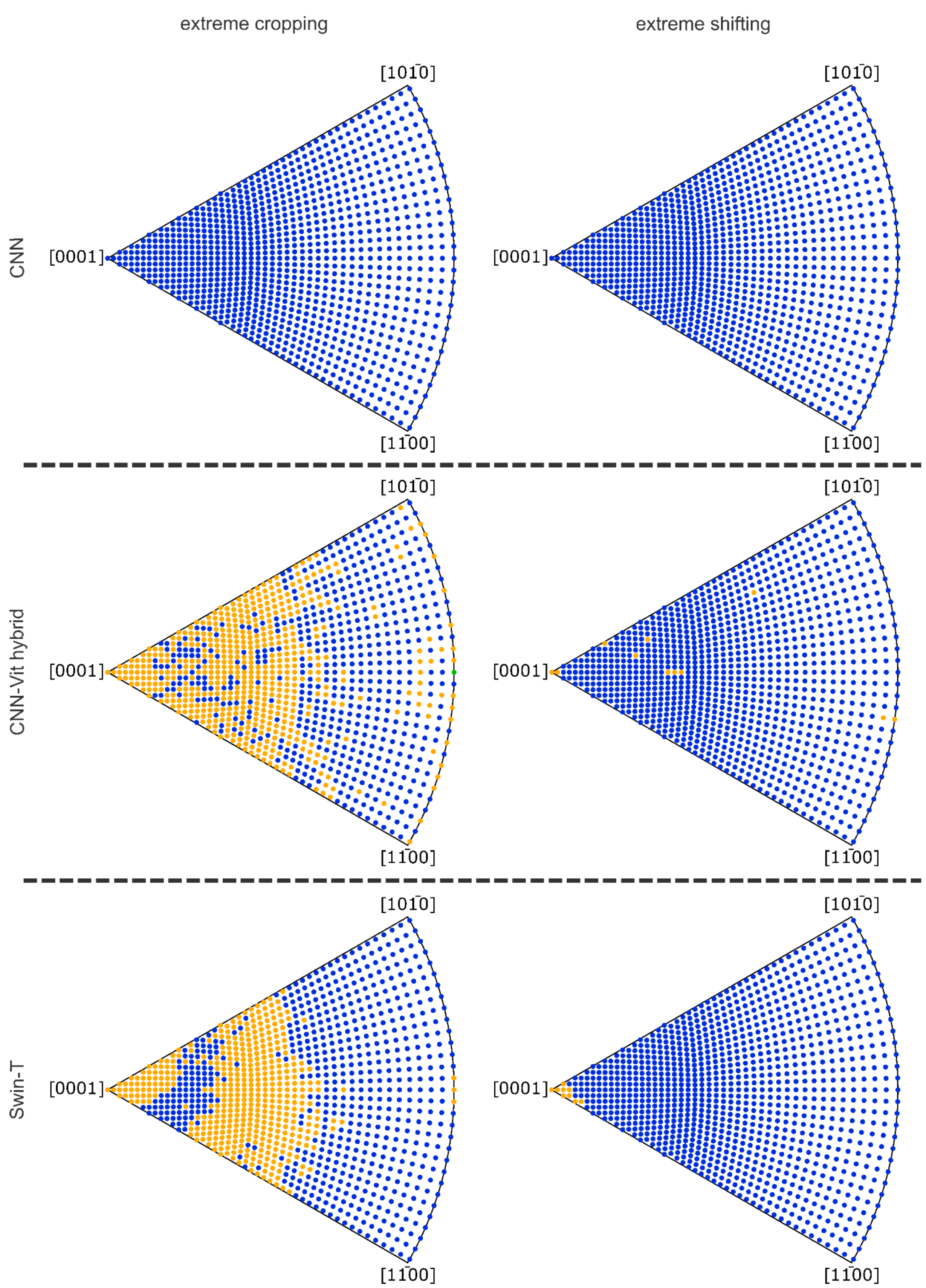

Fig 3: **Network predictions for extreme pattern shifts and crops of LNO diffraction patterns.**

Top: CNN; Middle: CNN-ViT hybrid model; Bottom: Swin-T.; Left: highly cropped image to simulate a strong deviation from the expected camera length. Right: strongly shifted diffraction pattern to simulate a significant descan. Blue: correct LNO prediction, Orange: false NiO prediction.

Since the [0001]-orientation of LNO seems to be the most problematic for the networks to differentiate from NiO, we let the NiO-orientation network predict the diffraction pattern of this orientation. The network predicts the Euler angle class [30°, 54.74°, 45°] for this pattern, which corresponds to the [111] orientation for NiO. In Figure 4 atomic models of the unit cells of LNO and NiO in these respective orientations are shown. It becomes clear that these orientations are equivalent to each other, and only differ in the composition of the atoms in the sublattice. For the NiO the sublattice consists solely of Ni atoms, while for LNO these are to some extent replaced by Li atoms. This equivalence of the orientation explains why the diffraction patterns are difficult for the networks to differentiate.

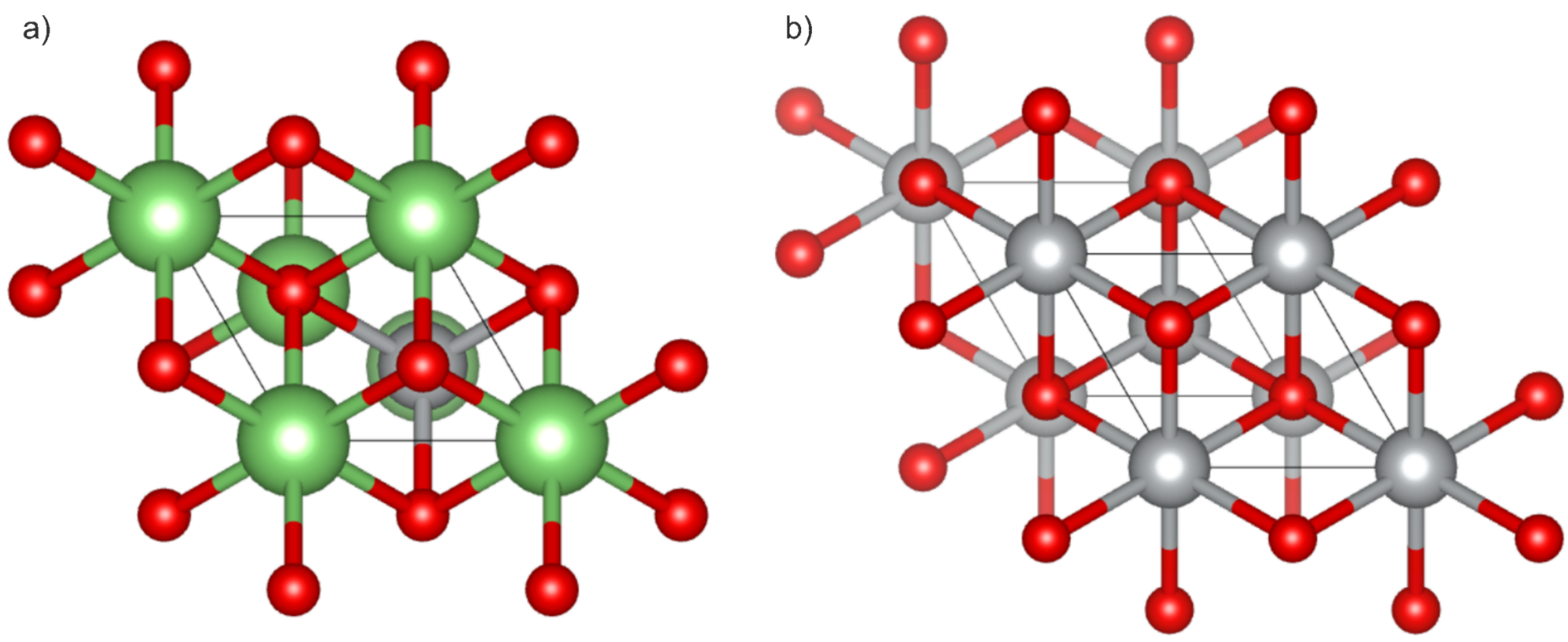

Fig. 4: **Atomic models of unit cells in problematic orientations.** a) LNO unit cell in [0001] orientation. b) NiO unit cell in [111] orientation. Green: Li-atom, Grey: Ni-atom, Red: O-atom.

We also test our networks by providing them with diffraction patterns of orientations not included in the original training dataset. For this, we identify particularly challenging orientations of both types of material for the networks and vary the Euler angles PHI and phi2 in steps of 0.2° from -2 up to 2° around those directions. Additionally, we test these images with varying scaling of the contrast in order to check how much post-processing might be necessary by a user to achieve optimal results. The predictions of our networks are shown in the Supplementary Figure S4. The results mirror our previous conclusions. The CNN and Swin-Transformer are again the most successful architectures, as they are able to correctly classify almost all of the unknown orientations. Additionally, all networks show improvements in their predictions by increasing the contrast of the image. The CNN and the Swin-Transformer, in particular, are less susceptible to changes in the visibility of the diffraction spots, keeping their high performance even with low visibility. There are also improvements seen by the combination of the patterns and the convolution feature maps to create the input for the transformer block of the CNN-ViT hybrid networks, while the pure vision transformer network is, for most tested orientations, basically unable to predict these unknown diffraction patterns.

A final test using simulated images is conducted by providing the networks with images that have been altered through augmentations that are not present in the training dataset. We chose a strong shear of the diffraction patterns as well as the inclusion of lines resembling Kikuchi lines that are typically encountered when not precessing the electron beam during the data acquisition. The latter are implemented by including thin lines of random lengths, orientations, and intensities into the diffraction pattern. These augmentations are implemented on the challenging diffraction patterns of the previous test. Supplementary Figure S5 shows examples of diffraction patterns augmented in this manner. Both such augmentations are significantly more challenging for template-matching algorithms. The shear alters the spatial relation of the diffraction spots to each other directly, and the lines can obscure diffraction points or be interpreted as relevant features by an algorithm that only detects intensities but does not take the actual shape of intensities into account. Overlap of a Kikuchi line with a diffraction spot of a template would still cause a positive activation for a conventional template-matching algorithm. Our networks demonstrate that they are very capable of dealing with these augmentations. The CNN and Swin-Transformer achieve 98% accuracy when predicting these diffraction patterns.

**Predictions on experimental test data**

To get a gauge on the networks' abilities to make predictions in a real-life application, we let them predict the phases from the datasets acquired from two experimental samples. The confidence values for the two orientation networks when predicting these samples, which are compared to get the phase prediction, are shown in the Supplementary Figure S6. We notice that, for all three datasets, our networks struggle to differentiate between a vacuum region and regions of amorphous material. This is found to be greatly improved by shifting the contrast of the image so that the intensity of the background is slightly raised; however, this also interferes with their ability to differentiate between the crystal phases. The most likely explanation for this behaviour is the presence of an afterglow in the scintillator of the CMOS camera when passing through vacuum regions of the sample, together with subtle variations in smoothly varying intensities (in empty regions of the diffraction patterns) that are not properly corrected during flatfield corrections. The enduring intensities are then amplified by the logarithmic scaling, giving the impression of diffuse scattering of an amorphous region. As the main focus of our networks remains the discrimination between the LNO and NiO phases, and such an intensity shift does have a moderate to significant effect on the networks' ability in this matter, we present here the predictions on the non-corrected images.

The first analysed sample was created using a sintering process at 800 °C for one hour [17]. This leaves us with a basically pure LNO sample. The predictions for this sample, shown in Figure 5, make it immediately evident that comparing the confidences of the two orientation networks remains challenging. Here we see that many of the individual grains of the sample are actually predicted to belong to the rock-salt phase, indicating that certain orientations are more difficult to differentiate between the phases. In addition, all the grain boundaries around the grains are also predicted falsely. The pure phase networks, on the other hand, all predict the sample regions almost entirely to be LNO. The CNN is the most successful in this regard, with basically no erroneous NiO prediction in the sample region. All the vision transformer models predict at least some presence of rock salt at some of the sample's edges. The non-hybrid ViT additionally predicts some of the amorphous region surrounding the sample as LNO, which obscures the geometry of the sample in our visualisation. Furthermore, the prediction of the transformer models for the non-sample regions is rather noisy, with many individual scan points predicted to belong to a crystalline phase. The CNNs, both the pure phase as well as the orientation networks, manage to produce a much cleaner prediction in this regard.

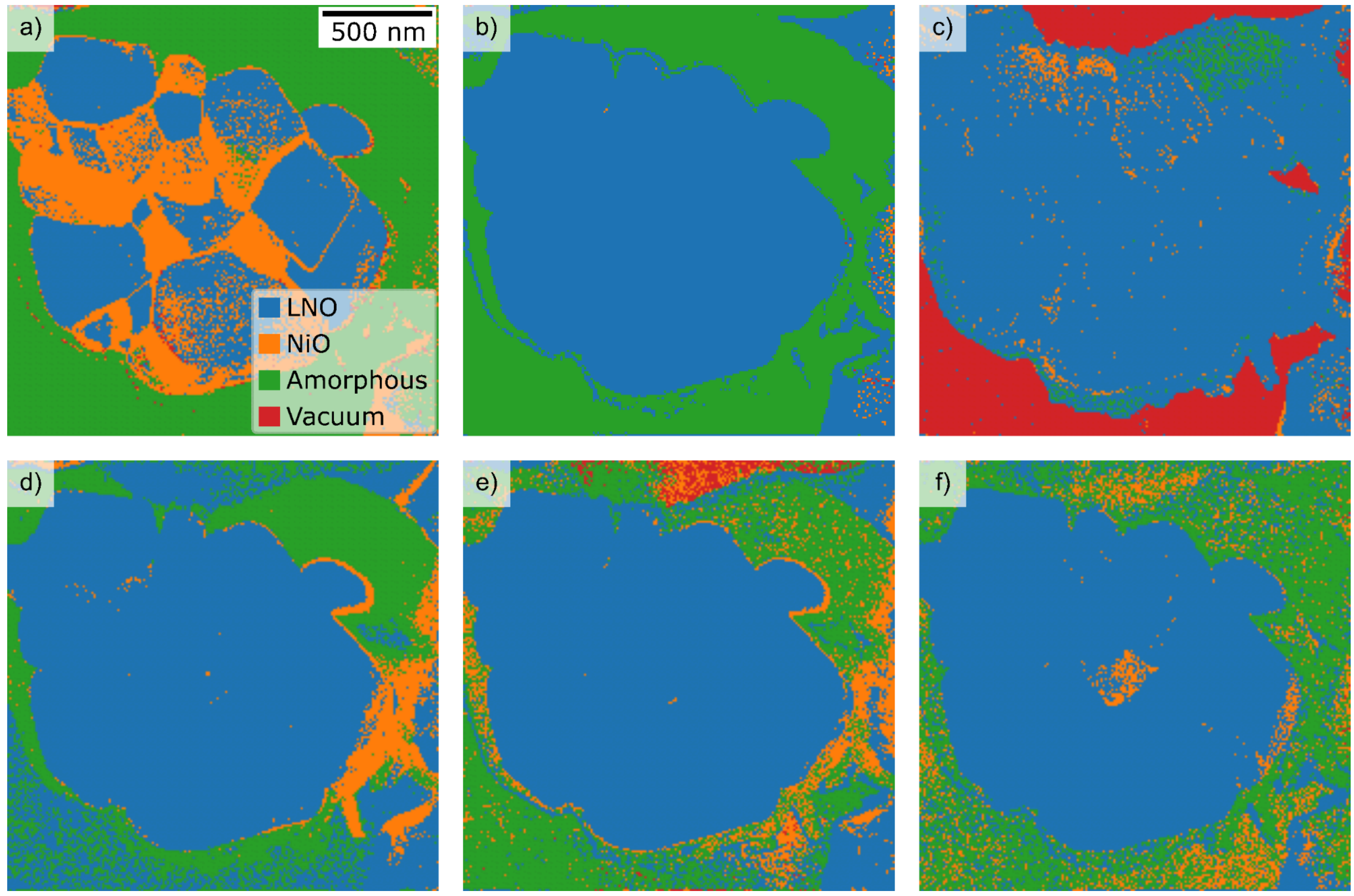

Figure 5: **Phase prediction of the networks on the pure LiNiO2 sample.** a) Confidence comparison of the two orientation CNNs, b) CNN, c) ViT, d) CNN-ViT hybrid network, e) CNN-ViT hybrid network with the original image as additional input for the CNN, f) Swin-transformer.

The second analysed experimental sample was created with the same method as the first. However, this is then followed with an annealing step at 700 °C for six hours [17]. This has proven to create a rock salt phase at the outer edge of the sample as can be seen in the results of a pattern matching phase analysis in the Supplementary Figure S7. Measurements using electron energy loss spectroscopy (EELS) show that this edge region consists in many areas of transitional phases between the two pure crystal types. Pure NiO appears actually quite rarely in this sample. We expect these areas to be difficult to predict for the networks, as they have only been trained with pure phases. The predictions for the different networks are shown in Figure 6. Comparing the confidences of the orientation networks, we again notice that all of the grain boundaries, internal as well as external, are classified as rock salt. In addition, significantly large areas inside the grains are predicted falsely. On the other hand, the phase networks all predict a thin rock-salt phase at the outer edge of the sample, in line with our expectation. The precise location and thickness of this layer varies from network to network, but it is centered around the middle grain of the three grains present. The pure ViT network also misclassifies the very top grain as belonging to NiO. Of the tested architectures, the classical CNN and the Swin-transformer appear to achieve the most accurate as well as cleanest results, with very few internal scan points misclassified.

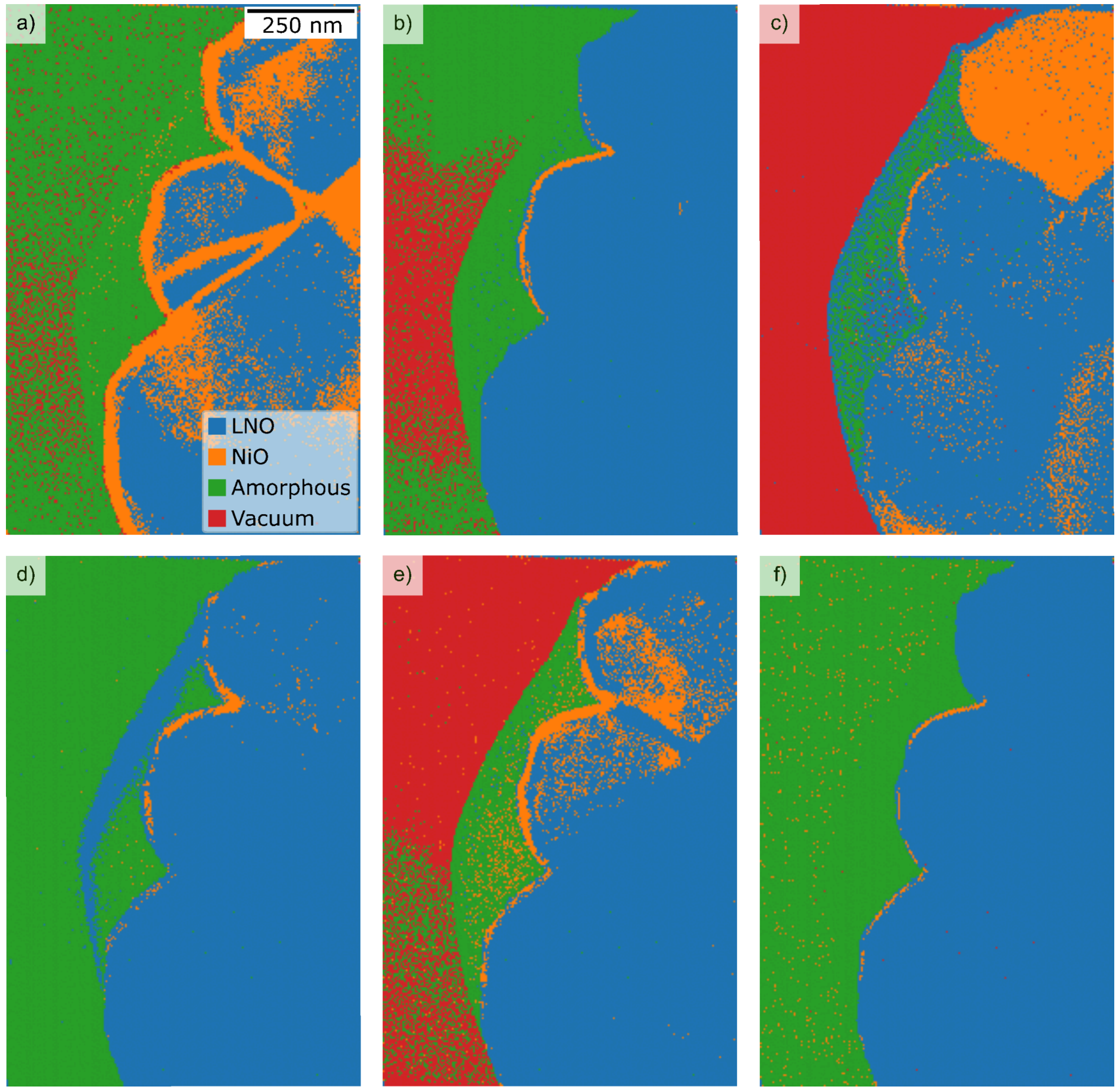

Figure 6: **Phase prediction of the networks on the annotated LiNiO2 sample.** a) Confidence comparison of the two orientation CNNs, b) CNN, c) ViT, d) CNN-ViT hybrid network, e) CNN-ViT hybrid network with the original image as additional input for the CNN, f) Swin-transformer.

The continued result of misclassifying the grain boundaries as rock salt, which we get from comparing the two orientation networks, would seem to suggest that, despite our best efforts, there still exists a clear bias towards this phase. This bias would show up especially in regions that are more challenging to predict, like grain boundaries. Though our results show that the method does work in principle, more work has to be put in in order to remove such biases.

For a third test dataset we use a collection of NiO diffraction patterns from three separate recordings. As these datasets feature a lot of overlapping grains, we use only a handful of patterns of non-overlapping areas. Exemplary diffraction patterns of this dataset can be found in the Supplementary Figure S8. Our CNN exhibits the most difficulties with this dataset. Only 60% of the patterns are predicted correctly. The ViT actually predicts every pattern correctly to be NiO. However, due to its poor performance with even the training data, we attribute these results to internal biases, rather than a proper prediction. The first CNN-ViT hybrid network performs better than the pure CNN, with 80% of the images correctly classified. The adaptation of adding the original image to the transformer block input raises this success further, with 100% of the patterns correctly predicted. The Swin-T is not quite able to replicate those results, with again 80% of the selected pattern predicted correctly.

**Analysis of the networks' decision-making for challenging cases**

Our results demonstrate that the networks are very capable in determining the phase for the vast amount of orientations. However, for a very few cases even the most successful networks start to struggle. In order to get a better understanding of the decision-making of our networks when predicting the phases of such challenging diffraction patterns, we implement the GradCAM method. This method creates heatmaps for images predicted by a classification network for each of the network's possible output classes. The heatmaps indicate how responsible different areas in the image are for the network when classifying it to a specific label. We limit our analysis to the classical CNN as it has shown to be extremely successful in this task, and its relatively simple architecture lends itself quite handily for such a method. Since our previous analysis shows that the [0001]-orientation of LNO is especially difficult for the networks to correctly classify, likely due to the strong similarity to a NiO [111] pattern, we provide synthetic images of these patterns at a thickness of 70nm to our network for analysis. To provide the networks with as much information and to keep the analysis as readable as possible, unaugmented diffraction patterns were used for this process.

Figure 7 shows the GradCAM heatmaps created by the conventional phase CNN for both classes overlaid on the LNO-input-images. When classifying the diffraction pattern as LNO, the network is only really influenced by one very specific area. This area is tracked by the network when the diffraction pattern is shifted around, as can be seen in Figure 7 b). Importantly, this area does not appear to contain any high-intensity diffraction spots. Only if this area falls to the very edge or even outside of the image does the network shift its attention to another region (Figure 7 c). This time, however, the relevant area is very clearly one specific diffraction spot. When classifying the same diffraction pattern as NiO, the region of interest is much more evenly spread across the whole image, with very clear gaps in the region around those spots that are relevant for a correct LNO prediction. The area most important for such a wrong NiO prediction seems to be the spots in the very center of the diffraction pattern. We also notice an area of particular unimportance for a NiO prediction in the form of a ring, which separates the relevant region into an inner and outer area. This appears to be a result of the gap between two Laue zones. Strikingly, the initial relevant area for an LNO prediction falls directly into this formed gap.

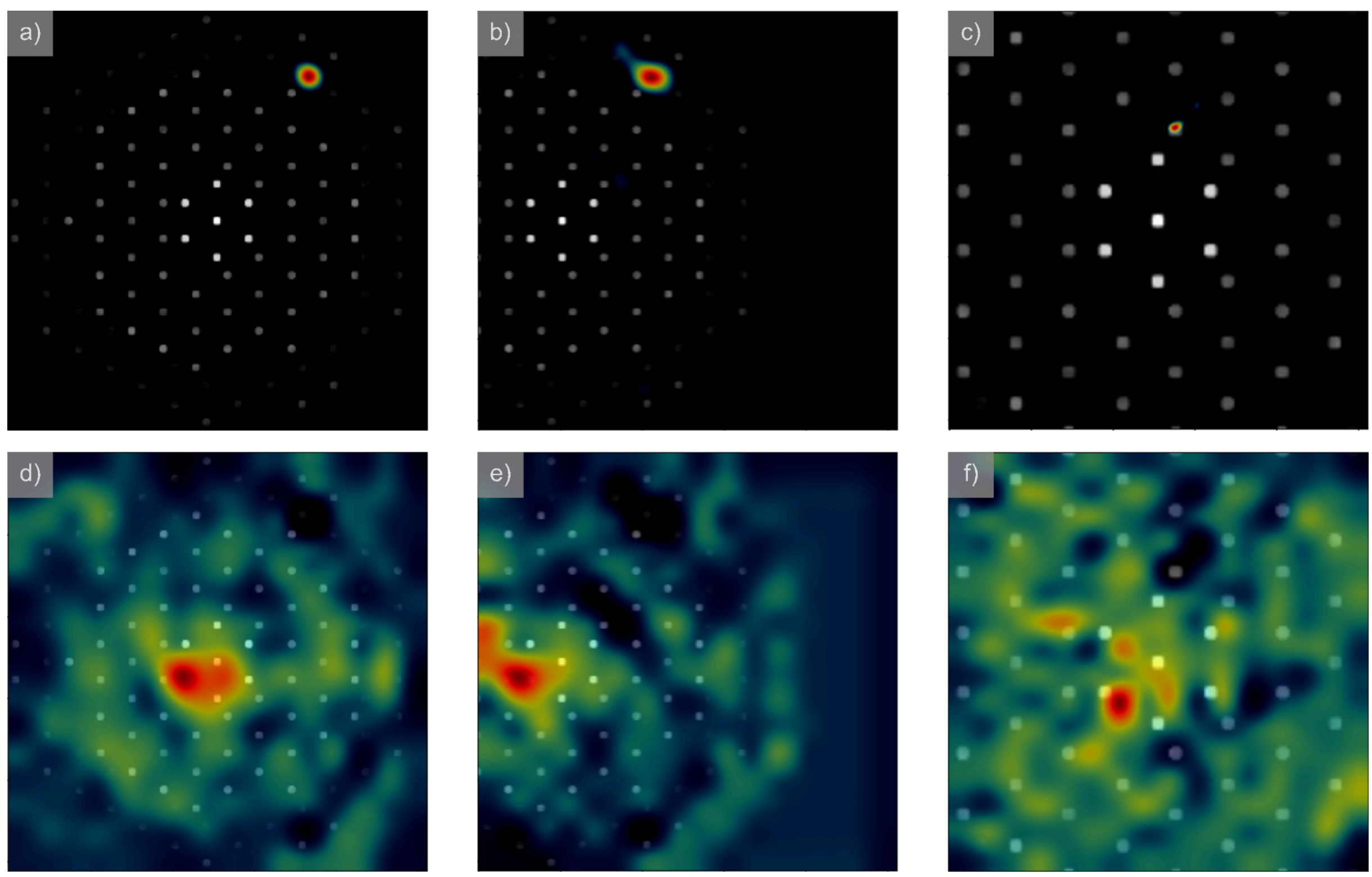

Figure 7: **GradCAM analysis of LNO pattern.** Heatmaps created by an implementation of the GradCAM method on the phase-CNN for a simulated [0001] LNO diffraction pattern. Top: GradCAM overlay for an LNO classification; Bottom: GradCAM overlay for a NiO classification.

In order to gain a deeper understanding of a wrong NiO prediction, we use the GradCAM analysis on the NiO diffraction pattern discussed above. Figure 8 shows GradCAM analyses on the NiO diffraction pattern for both a prediction of LNO and NiO by the conventional phase-CNN. In the case of an LNO prediction, we again observe that only very specific areas are responsible for the prediction. We note that one such region in the lower right of the image fits a specific diffraction spot. However, another region is roughly in the same area as the important area in the previous analysis. This region surrounds a diffraction spot very closely without overlapping it. When we shift and/or crop the image, we detect this even more clearly. The network is carefully avoiding the diffraction spots and focusing on seemingly empty regions. For a now correct NiO prediction, we see again the evenly spread out attention with a clear gap separating the heatmap into an outer and inner region.

A direct comparison of the two similar diffraction patterns, [0001]-LNO and [111]-NiO, with a power scaling typical for our training data can already provide valuable insight (Figure 9). Only seven very central spots of the two diffraction patterns are located exactly on top of each other and have significant intensity. The previously discussed focus of the phase network on these central spots, when falsely predicting LNO as NiO, can be explained by this. In Figure 9 b) we overlay the two relevant diffraction patterns at a thickness of 70 nm and set the intensity to a binary scale so that all diffraction spots are equally visible. We mark the most relevant areas for an LNO with dotted lines. It becomes clear that the tracked area in the LNO image (marked in purple) does, in fact, contain diffraction spots in both patterns, which do overlap almost perfectly. Though it is a spot of lower intensity in both patterns, it is far more visible in the LNO pattern. This might be the reason why the network uses it as an indication of the phase. The important regions in the NiO image (marked in green) are also shown to contain spots. In these cases, the network tries to avoid spots on the NiO pattern to only check LNO spots, which are very close to NiO spots. In those cases, we also note that the relevant spots fall inside or very close to a spotless ring in the NiO pattern, which forms the border between the zeroth and first order Laue zones. This absence of NiO spots would make it easier to categorise a spot as belonging to an LNO pattern, which might explain the network's decision to look for spots here. It is also the reason for the gap in the heatmap for any NiO prediction, as less general intensity in this area is expected for a NiO pattern.

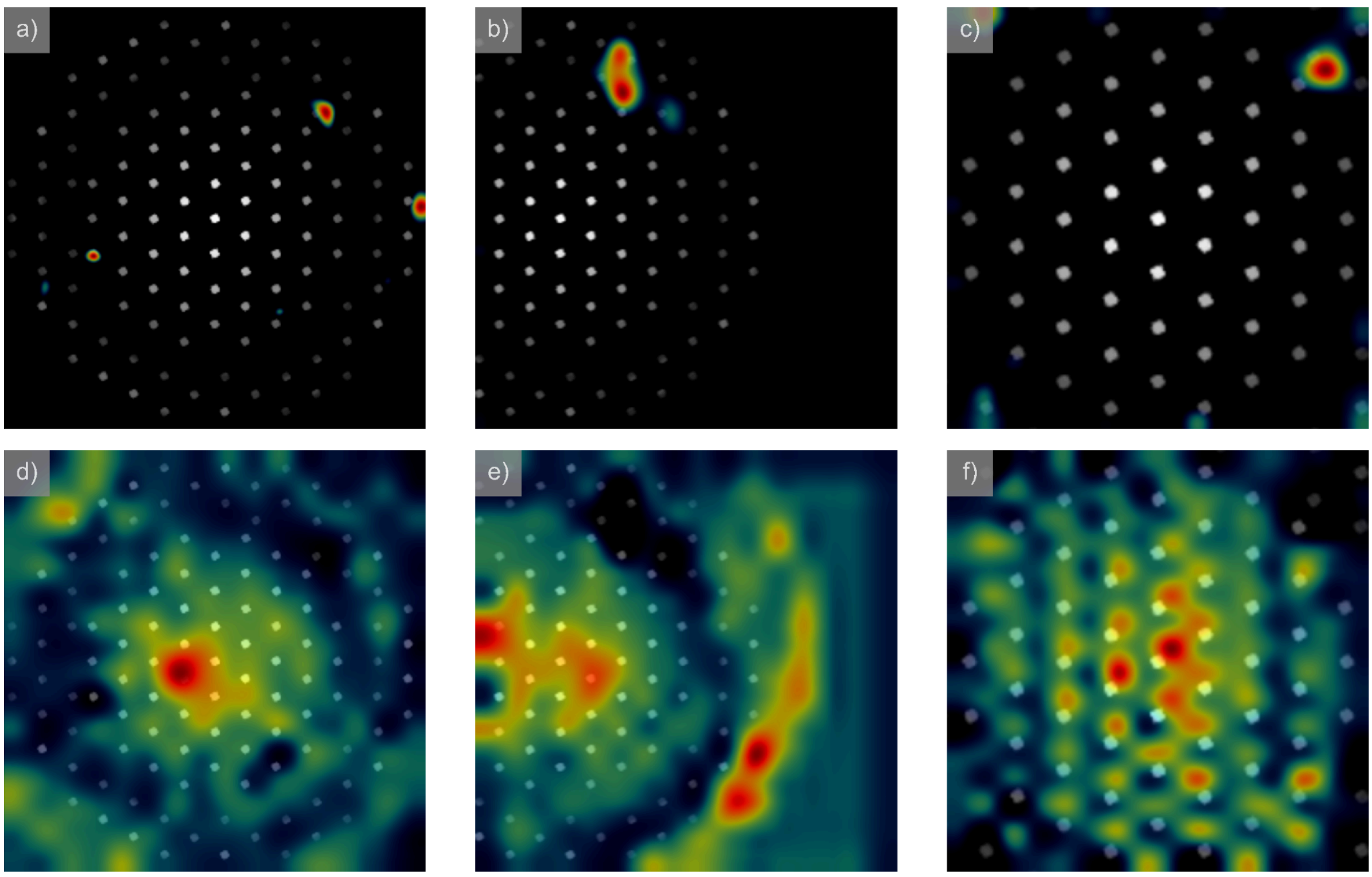

Figure 8: **GradCAM analysis of NiO pattern.** Heatmaps created by an implementation of the GradCAM method on the phase-CNN for a simulated [111] NiO diffraction pattern. Top: GradCAM overlay for an LNO classification; Bottom: GradCAM overlay for a NiO classification.

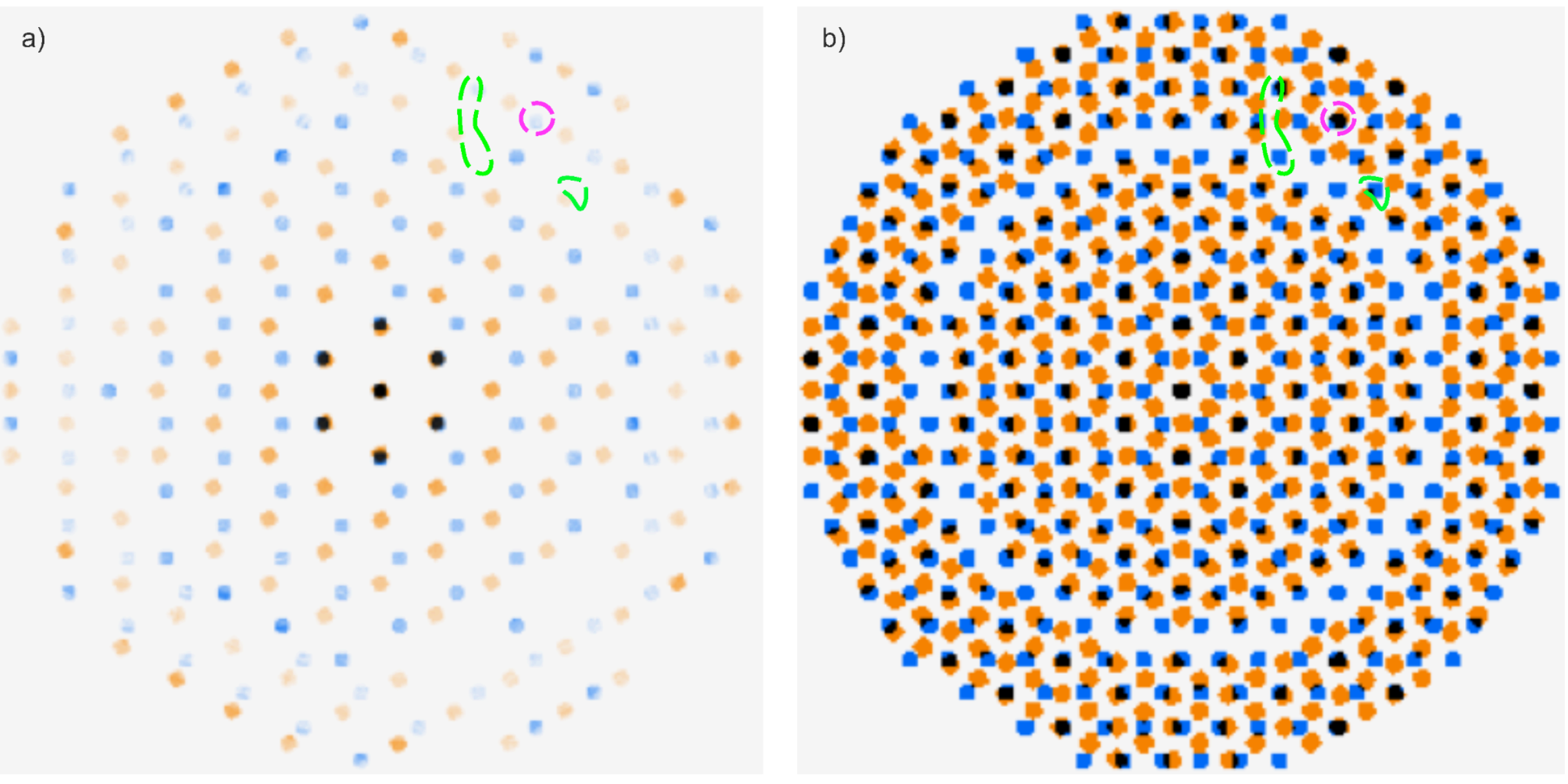

Figure 9: **Overlay of the diffraction patterns used for the GradCAM analysis.** Blue: LNO [0001], Orange: NiO [111]. a) diffraction spots scaled with original intensity, b) binary intensity to make every spot clearly visible. Areas marked with green and purple dotted lines represent areas of interest for an LNO prediction in the NiO and LNO patterns, respectively.

The gap between the Laue zones is formed by a threshold value in our simulations, which completely excludes spots entirely from the intensity calculation if they are too distant from the Ewald sphere. The exclusion of spots from the calculation can significantly alter the intensity of individual spots. Supplementary Figure S9 shows overlays of the two analysed patterns in which the threshold for the LNO pattern was decreased to display a similar gap as NiO. In this case, a larger number of spots with significant intensity of both patterns overlap almost perfectly. To review if the choice of this threshold value significantly changes the performance of our networks, we create an additional training dataset in which this parameter is varied and use it to retrain all network architectures. The results differ in many ways from our original networks. For one, even the CNN is now fooled and wrongly predicts a NiO phase for simulated LNO diffraction patterns very close to the [0001]-orientation. When analysing the second experimental LNO sample, these networks display very noisy predictions inside the LNO grains and all except the CNN are unable to predict a NiO boundary on the outer sample edge. And the CNN now predicts one of the grains to be wholly of the rock-salt phase. On the other hand, all phase networks trained with this new data are predicting the third experimental dataset of purely NiO patterns perfectly. Results for a classical CNN trained with those images can be found in Supplementary Figure S10. More work is needed to investigate if the predictions of a network trained with more varied data can be stabilised.

## Discussion

In this work, we investigated multiple network architectures for their applicability to phase mapping. We found that the classical convolutional architecture as well as the Swin-Transformer architecture achieved the overall best results. Simple Vision-Transformers failed at this task quite drastically and only improved by including a convolutional block to preprocess the input pattern. Providing the vision transformer with both the original input image as well as the output of the convolutional block as a combined input further improves the predictions of the ViT in most cases. Our analysis showed that the very practical approach of comparing the confidences of individual orientation networks can lead to quite usable phase predictions. However, when confronted with patterns of challenging areas, like grain boundaries, this method exhibits a tendency to favour one phase over the other. More work is needed for this method to fully rival the predictions of dedicated phase networks. We also found that the [0001]-orientation of LNO in particular seems to be a challenge to our networks due to its pattern's similarity to the pattern of the [111] NiO orientation. A GradCAM analysis of the decision-making of the most successful network architecture highlights a peculiar behaviour for this orientation. Only a very few low-intensity spots appeared to be relevant for a LiNiO2 prediction. Blocking such individual spots during the training might be a possible approach for future work in order to challenge the networks further and force them to learn a more stable prediction approach.

To improve prediction results in subsequent works, we envision multiple paths. For one, the synthetic training data could be improved by including more effects and therefore more information for the networks to analyse. Combining the diffraction patterns with an accurate

Kikuchi band simulation or switching the random noise to a more accurate but computationally expensive Poisson noise might provide such avenues. A second idea is to further widen the variability of the training data. For this more simple changes in the simulation, like a varied beam convergence angle could be implemented. Additionally, the patterns could be augmented to reflect different detector types. Furthermore, the training dataset could be expanded by the inclusion of experimentally recorded diffraction patterns. Though, since 4DSTEM recordings tend to generate large datasets of only a handful of different orientations, great care would have to be taken to make such images in the dataset frequent enough to be significant, while at the same time not create biases towards certain orientations. Finally, the training data might be improved by the inclusion of overlapping patterns. These tend to exist at the very boundaries of sample grains and might be the cause for the poorer performance of the orientation network approach in our results.

## Methods

### Training data generation

The diffraction pattern of a TEM sample changes substantially depending on the orientation of the sample in the electron microscope. Therefore, a network needs to receive a training dataset that encompasses as much of the potential orientation space as possible to learn to differentiate between different phases. Creating such a complete dataset with experimentally acquired images is nearly impossible for most materials. Instead, we decide to use synthetic training data by simulating diffraction patterns. This provides a way to generate a large amount of highly varying data with perfect labeling in a manageable timescale. We utilise the py4DSTEM package for Python, which employs a Bloch-wave algorithm to simulate the diffraction patterns of both LNO and NiO. The angular steps between the simulated orientations are determined in an equidistant measure with the help of the "orix" package for Python [18]. This way of filling the angular space is the most effective for neural networks to learn to differentiate the sample orientations [19]. This method only allows for variation of two of the three Euler angles needed to define an orientation (PHI, phi2) in the Bunge notation. Variation in the last angle (phi1) is achieved by a simple in-plane rotation of the resulting diffraction pattern. Because the two phases possess different crystal symmetries, the number of simulated orientations differs strongly. Therefore, separate resolutions of the orientation space are chosen. For LNO the resolution is 2° while for NiO it is 1°. The number of different simulated orientations for LNO is 152,120, while the number for NiO is 38,918. Figure 10 shows a visualisation of the simulated orientations for both materials without the in-plane rotation. Since the intensities of the diffraction spots in the patterns shift with varying sample thickness, all images are simulated for four different thicknesses, i.e., 10, 30, 50 and 70 nm.

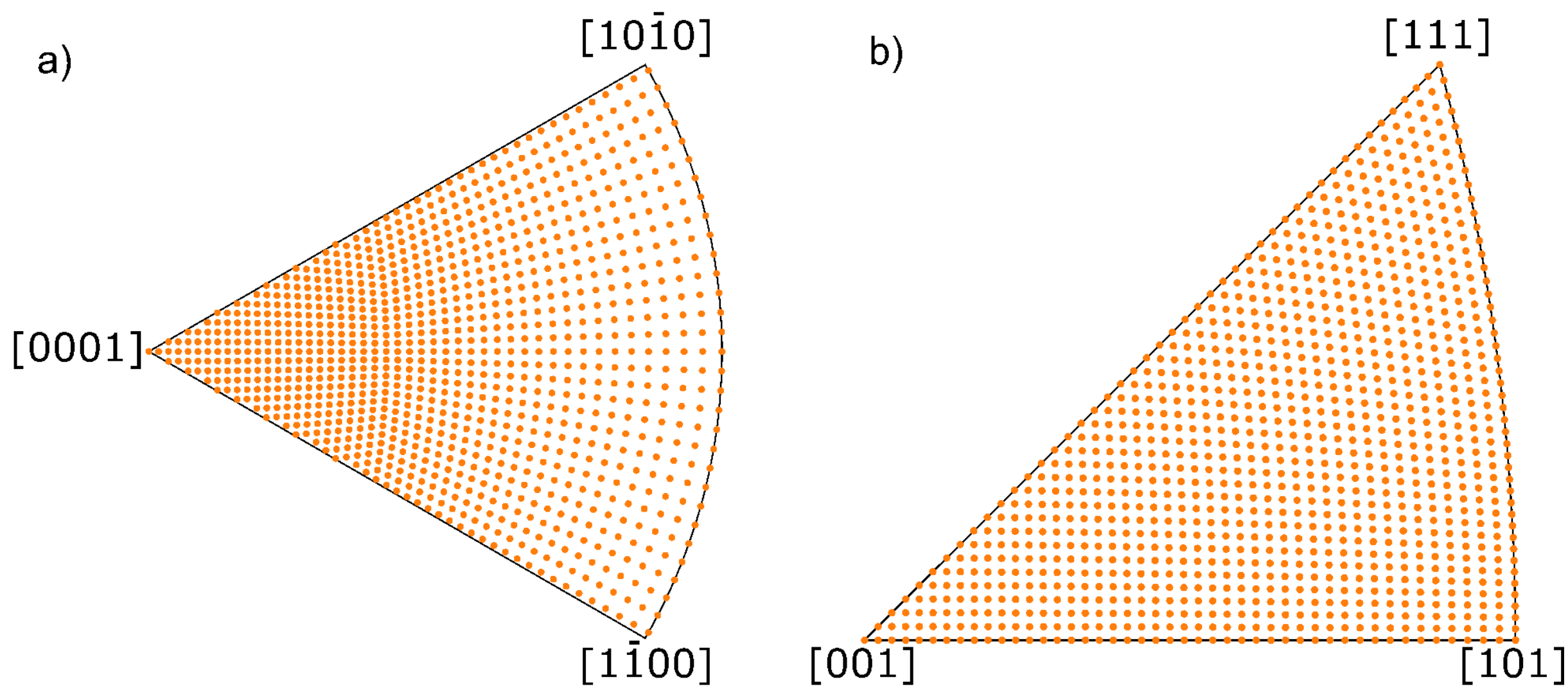

Figure 10: **Visualisation of the orientations in the training dataset.** a) LNO and b) NiO. Variation of the phi1 angle is not included.

To increase the number of training images and make the networks robust against obstructive effects of the electron microscope, we employ several augmentations on our simulations. Since noise is always a factor in TEM images, we include it in the form of random noise. To include diffuse scattering, we add intensity in the form of a Gaussian bell. Moreover, the effect of amorphous material on the surfaces of the sample is included by the addition of a superposition of sine and cosine intensity waves to the diffraction patterns. The shapes and intensities of the bell and waves are randomised and only added to half of the images. After that, the diffraction patterns are randomly cropped to train the network for different camera lengths and misalignment of the diffraction pattern during the experiment. Then, the maximum intensity of the pattern is randomly cut to simulate different clipping values of the camera systems. Finally, we utilise a Gaussian filter to smooth out the resulting image. In total, 30 differently augmented patterns were created for LNO for each simulated orientation. For NiO this number was raised to 120. This ensures that a similar number of images are present for both materials, which reduces the chance of the network internalising a bias towards one phase. Figure 11 shows a simulated diffraction pattern of LNO before and after the augmentations.

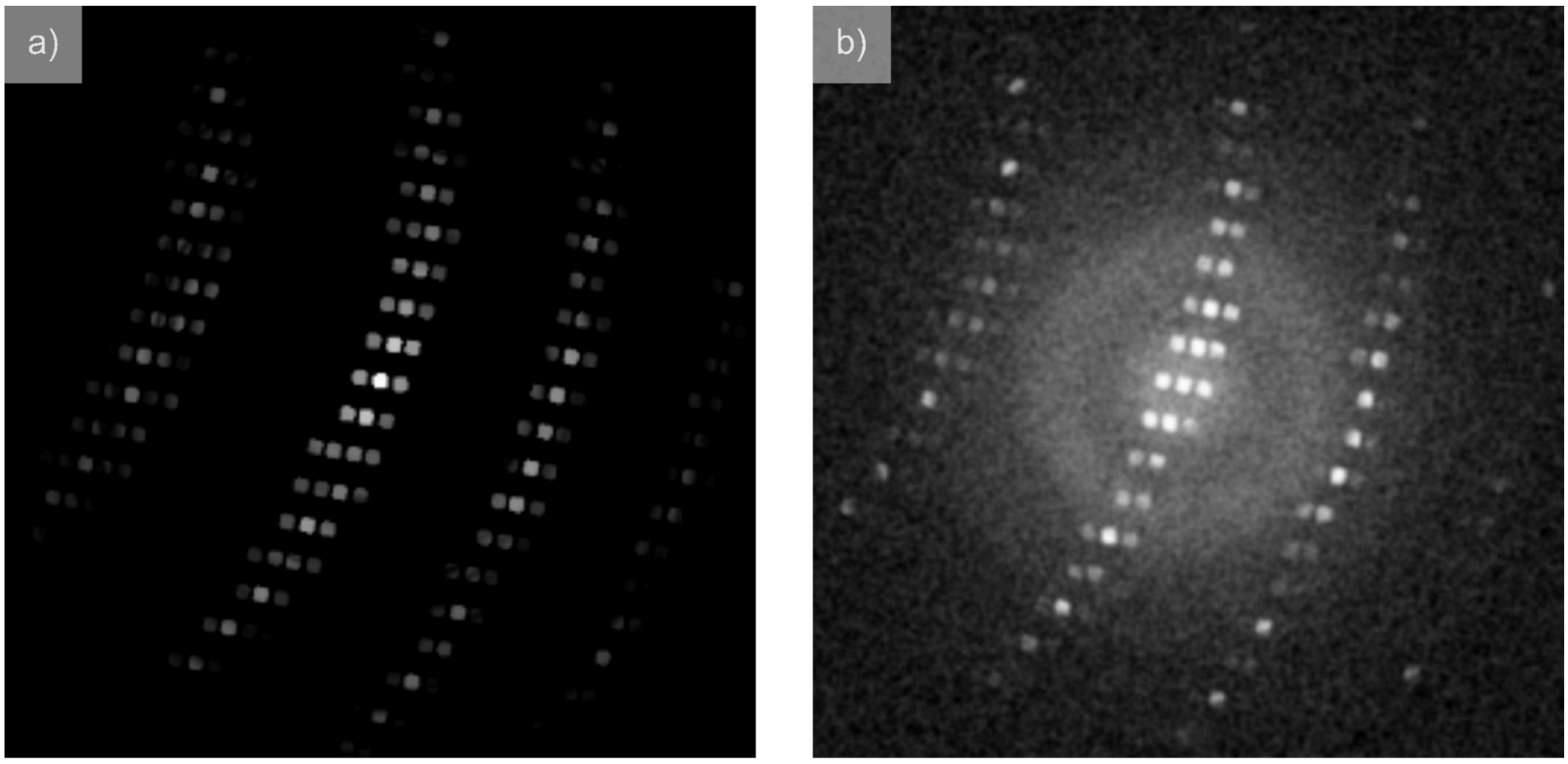

Fig 11: **Example of implemented augmentations.** a) Unaugmented diffraction pattern of LNO simulated with a Bloch-wave algorithm. b) Diffraction pattern augmented by adding random noise, diffused intensity and intensity rescaling.

To create diffraction patterns for the amorphous phase, an image of a single direct beam is simulated. This image is then augmented in the same way as the images of the crystal phases. To create images for the vacuum phase, the ring intensity and diffuse scattering augmentations are not applied to the zero-thickness image. For each of those phases, 450,000 images are created. Putting it all together, the final dataset consists of around 3.7 × 10^6 patterns.

**Test data generation**

In order to test the abilities of the networks, a mixture of simulated and experimentally recorded datasets is used. First, multiple orientations are identified for which the diffraction patterns of LNO and NiO look very similar, e. g., LNO in the [2 1 0] and NiO in the [1 1 2] orientation. For these, additional diffraction patterns are simulated by varying the PHI and

phi2 angle by steps of only 0.1 mrad up to 2 mrad. With this dataset, we test if the networks are able to differentiate between the phases for orientations they have not been trained on. Additionally, we create another dataset from these simulations by adding extremely severe augmentations, including a strong shear of the image. With this dataset, we test if the networks are able to handle highly unfavorable imaging conditions.

For experimentally recorded data, three different datasets are used: The first dataset consists of recordings of a pure LNO sample with many grains of various orientations. It was synthesised using a sintering method at 800 °C for one hour. The second dataset is also acquired from a cross-section of an LNO sample. However, through an annealing process at 700 °C for six more hours , the phase at the very edge of the sample has been transformed into various mixtures of the LNO and NiO phases [17]. Figure 12 shows virtual dark field overview images of these two experimental samples. In order to verify that our networks are predicting the rock-salt phase correctly in experimental situations, we also need a pure NiO dataset. However, our samples feature often overlapping grains. This changes the resulting diffraction patterns significantly and presents a challenge that our networks are not trained for. Therefore, we form our test datasets by collecting individual diffraction patterns from selected areas of non-overlapping grains from three different NiO samples. Training networks to be able to handle overlapping grains would be an intriguing prospect for future work.

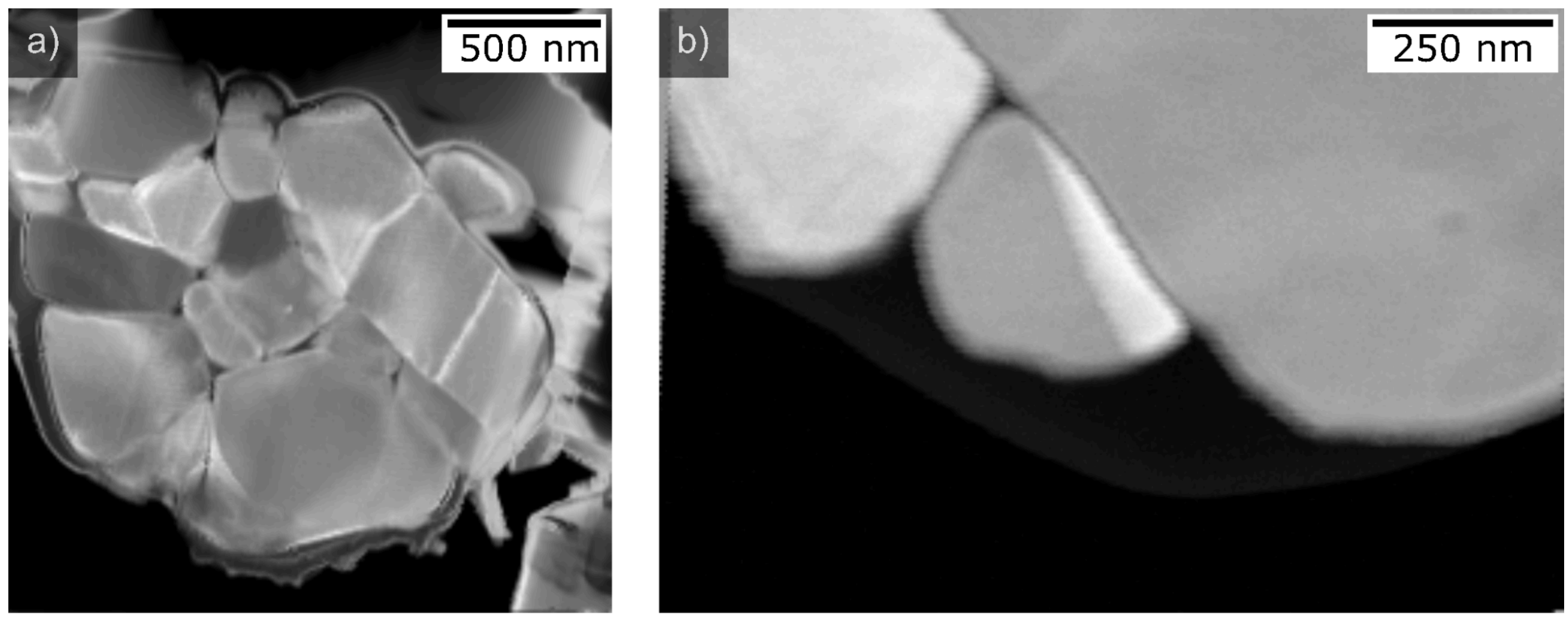

Fig 12: **Experimental validation samples.** Virtual dark-field overviews of the two experimental LNO samples used to validate the networks. a) pure LNO sample, b) LNO sample with a rock-salt phase at the outer edge induced by annealing.

## Network architectures and training methods

In this work, we compare results from multiple network architectures. To begin with, we employ classic convolutional neural networks (CNNs) for their simplicity and proven record when applied to TEM data classification. These networks use a series of layers of an increasing number of convolution filters combined with pooling layers to transform the input image into a small-dimensional latent space that contains the relevant information for the succeeding classification. Our convolutional blocks consist of 3x3 filters activated with a ReLU function and followed by batch-normalisation and 2x2 max pooling layers. The number of convolution filters is doubled with each block, starting with 8 until reaching 128. After these blocks, the final feature maps are flattened and fed into two fully connected layers of 3,200 nodes before reaching the output layer, which is activated with a softmax function. We also implement a dropout layer between each fully connected layer to reduce the chance of overfitting.

We further employ a vision transformer (ViT) architecture. These networks split the input image into smaller tiles that are then embedded into a vector together with a positional encoding. Those patches are then fed into a series of transformer blocks, which consist of a multi-head self-attention layer and a feed-forward block. The former calculates the relevance of the patches to each other for different issues, and the latter acts like a small CNN. In this study, the images are split into patches of 16x16 pixels, which are subsequently projected into a 64-dimensional space and then fed through 6 transformer blocks. Each of those blocks consists of an attention layer with 12 attention heads and two fully connected layers containing 128 and 64 nodes, respectively. The transformer blocks are then followed by two fully connected layers with 1,600 nodes each and a final output layer. All fully connected layers in the ViT architecture are activated with a GeLU function [20].

We also create a CNN-ViT hybrid model to combine the strength of the convolutional network to identify local patterns with the transformer's ability to identify long-range dependencies between different image regions. Here, a series of three Convolutional layers is put before the ViT structure. The number of filters is doubled at each layer, starting with 32, and is also activated with a GeLU function. Differently from the classical CNN model, no max pooling [21] was utilised to keep the output of the convolution part at the same dimension as the input image. The resulting feature maps of the third convolutional layer are then used as input for the ViT structure.

This hybrid architecture is also further adapted into another model in which the original input image is combined with the final feature maps of the CNN structure as the input for the transformer network. In this way, the ViT part of the network is able to analyse the raw image data as well as the features extracted by the CNN part. Additionally, it is hoped that the ViT's strength at performing long-range analysis translates here to the network drawing correlations and conclusions between the features and the input image. The architecture for this network is shown in Figure 13.

Finally, we also implement a Swin-transformer network. This represents an adaptation of the Vision transformer that progressively merges the embedded patches into larger clusters. This allows the network to analyse both low and high-level features, while still being computationally manageable, similar to classical CNNs. To further limit the computational cost, Swin-transformers do not calculate a global attention between all patches but limit the attention to smaller non-overlapping areas that are shifted around. For our study, we use a SwinTransformerV2Tiny256-network from the Python package "tfswin" which is pretrained on the "ImageNet-1K" dataset [22]. The network is then adapted by removing the original output layer and replacing it with a global average pooling layer and a subsequent fully connected layer to fit the architecture to the correct number of output classes. As this network architecture expects RGB images as inputs, we also have to convert our grayscale training images for this network.

We train two types of models that differ in their respective output layers. For the first type, the output layer consists of only four nodes, one for each phase (LNO, NiO, amorphous, vacuum). For the second type, separate networks for both LNO and NiO are trained. These employ as many output nodes as there are different orientations of said material in the training dataset. The networks learn, therefore, to predict the orientation of the material directly. The phase can then be determined by comparing the confidences of the two networks when predicting the same input image. Other methods to compare which of the two networks' outputs should be chosen in order to assign a phase to an image were considered, but found to be less successful. In Supplementary Figure S11 confusion matrices utilising comparisons of the confidence, entropy, energy scale and the pure logits are shown. Given the extreme number of output nodes and with that the number of trainable parameters, we only utilise this method for the relatively small CNNs, as the transformer models are already very large and the training and subsequent prediction time would suffer greatly with this approach.

Our networks are trained over a period of between 600 and 1,200 epochs until no more improvements are achieved in the validation loss. Each epoch consists of 128 batches of 128 patterns. Each training epoch is followed by a validation epoch of 32 batches. We employ an adaptive learning rate that gets reduced by 50% if the validation loss is not diminishing after 20 consecutive training epochs. This reduces the step size the network takes on the multi-dimensional loss-function plane and reduces the chance of the network jumping around a localised minimum.

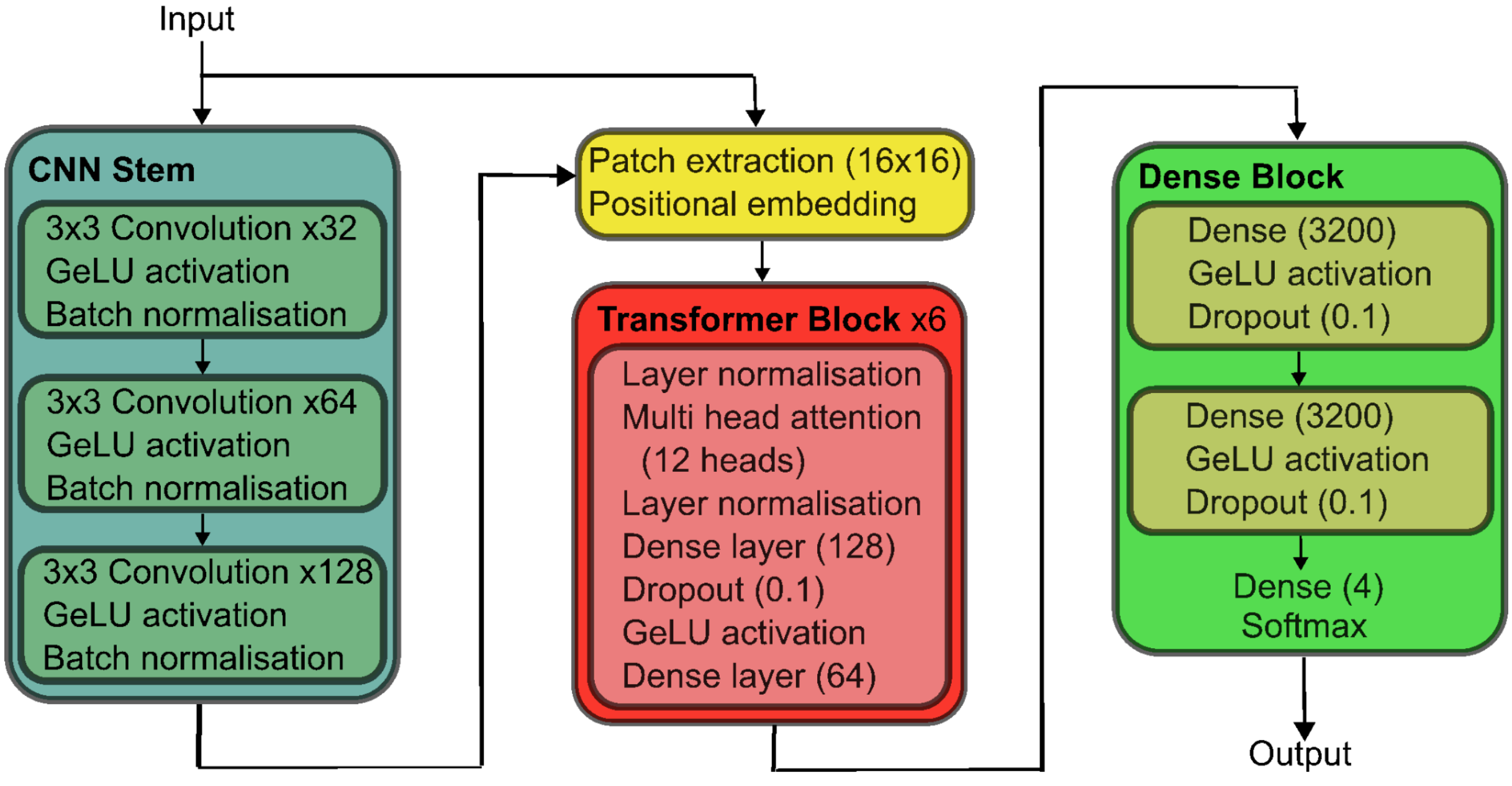

Fig 13: **Exemplary overview of the CNN-ViT hybrid architecture.** The input diffraction pattern is fed through a small convolutional block without pooling layers to keep the image size constant. The resulting feature maps of the last Conv-layer are subsequently combined with the original pattern and then transferred to the Transformer block. Its output is further processed by a block of two Dense layers before a final Dense layer of only four nodes is activated using the Softmax function to provide the output predictions.

**Network calibration and confidence scaling:**

The softmax layer at the very end of a classification network transforms the network's outputs into a pseudo probability, typically interpreted as the network's confidence, where the values given to all the classes add up to one. However, when we compare the predictions of multiple orientation networks to decide which phase an input image belongs to, outputs that more closely resemble actual probabilities are preferable. When plotting the networks' confidences against their actual accuracy (Figure 14), we see that our networks are very underconfident in their predictions. For this reason, we implement temperature scaling as a post-hoc calibration method for the orientation networks [23]. The temperature value is varied to find values that minimise the combined gap between the output confidences and the accuracies. The optimum temperature values found are 0.41 for LNO and 0.35 for NiO, respectively.

An additional challenge for confidence comparison arises from comparing different numbers of output nodes. A theoretical network which is trained to differentiate between only two classes could only ever produce a lowest confidence value for a given image of 50%. If one were to compare such a network to one with many thousands of classes, the confidence of the binary network could easily trump that of the second network, even if the latter is trained to handle the particular image and is able to categorise it, while the binary network is not. Therefore, we also rescale the outputs of our orientation networks in such a way that the value 1/(number of classes) is mapped to 0, while 1 remains the maximum value. This ensures that the two compared networks work with a similar baseline, as the particular confidence, which a network now assigns to an image that it can not fit to any of its classes, is now 0, irrespective of the number of classes the network trained for.

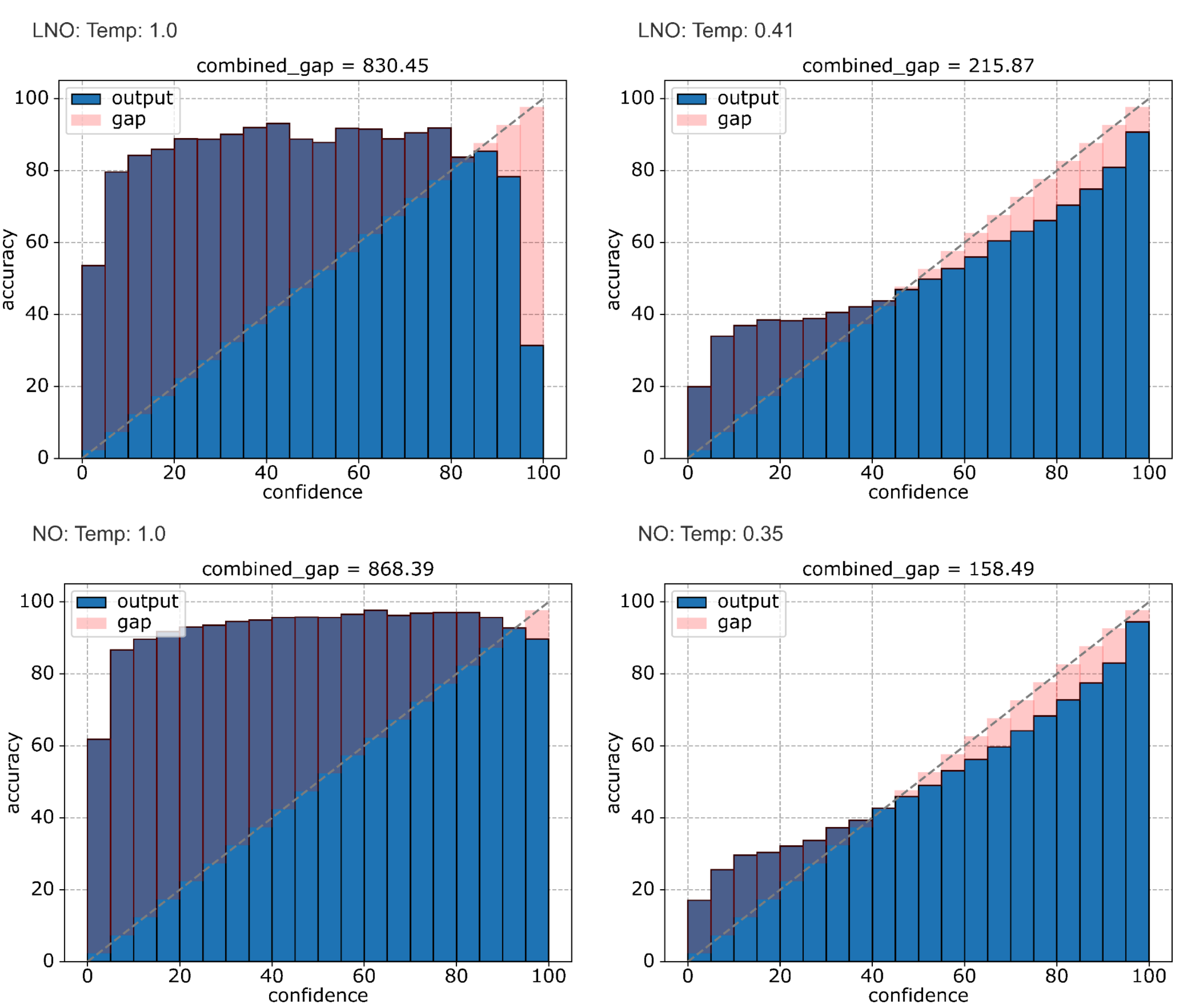

Fig 14: **Calibration of the orientation CNNs.** Comparison between the assigned confidences of the orientation networks and the actual accuracy with temperature scaling implemented.

Left: temperature 1.0, which is the same as no scaling. Right: temperature 0.41 for LNO and 0.35 for NiO.

## Competing interests

No competing interest is declared.

## Supplementary Information

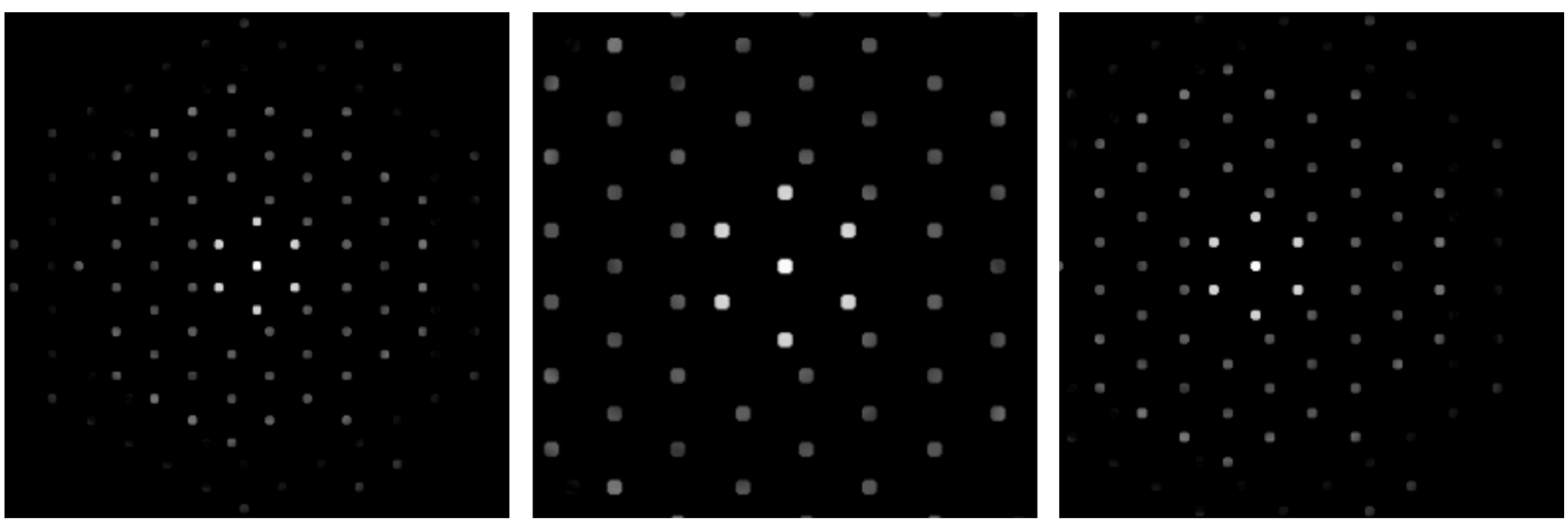

Fig S1: **Example of simulated user errors.** Comparison of an unaugmented diffraction pattern (a) to the extremely cropped (b) and shifted (c) patterns used to test the networks' stability to user error.

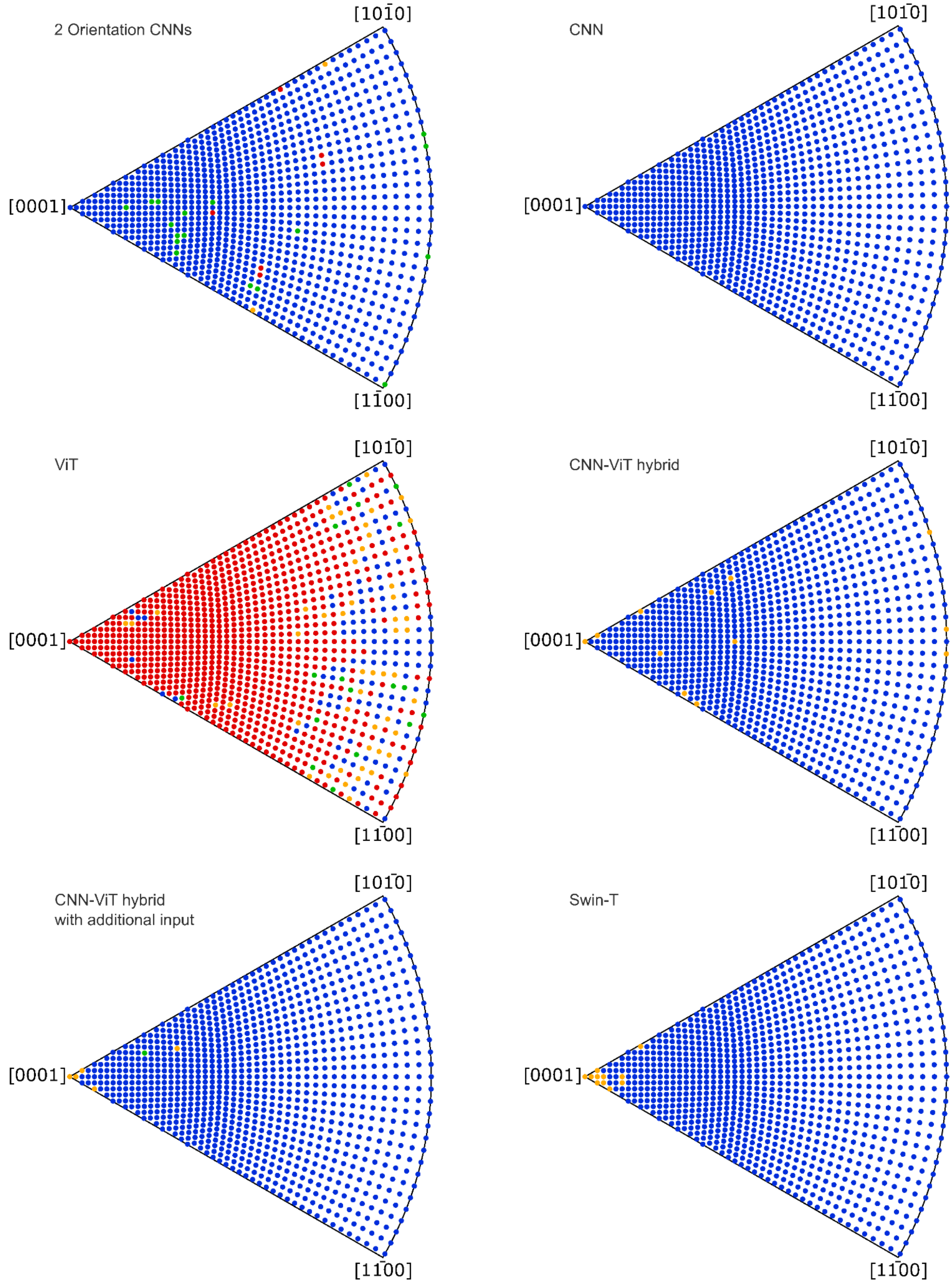

2 Orientation CNNs
[10$\bar{1}$0]
[0001]
[1$\bar{1}$00]
CNN
[10$\bar{1}$0]
[0001]
[1$\bar{1}$00]
ViT
[10$\bar{1}$0]
[0001]
[1$\bar{1}$00]
CNN-ViT hybrid
[10$\bar{1}$0]
[0001]
[1$\bar{1}$00]
CNN-ViT hybrid
with additional input
[10$\bar{1}$0]
[0001]
[1$\bar{1}$00]
Swin-T
[10$\bar{1}$0]
[0001]
[1$\bar{1}$00]

Fig S2: **Phase prediction on LNO data.** Visualisation of predictions made of the networks on non-augmented synthetic LNO images for all orientations in the training dataset (excluding phi1 variation). Blue: LNO, Green: NiO, Yellow: amorphous, Red: vacuum.

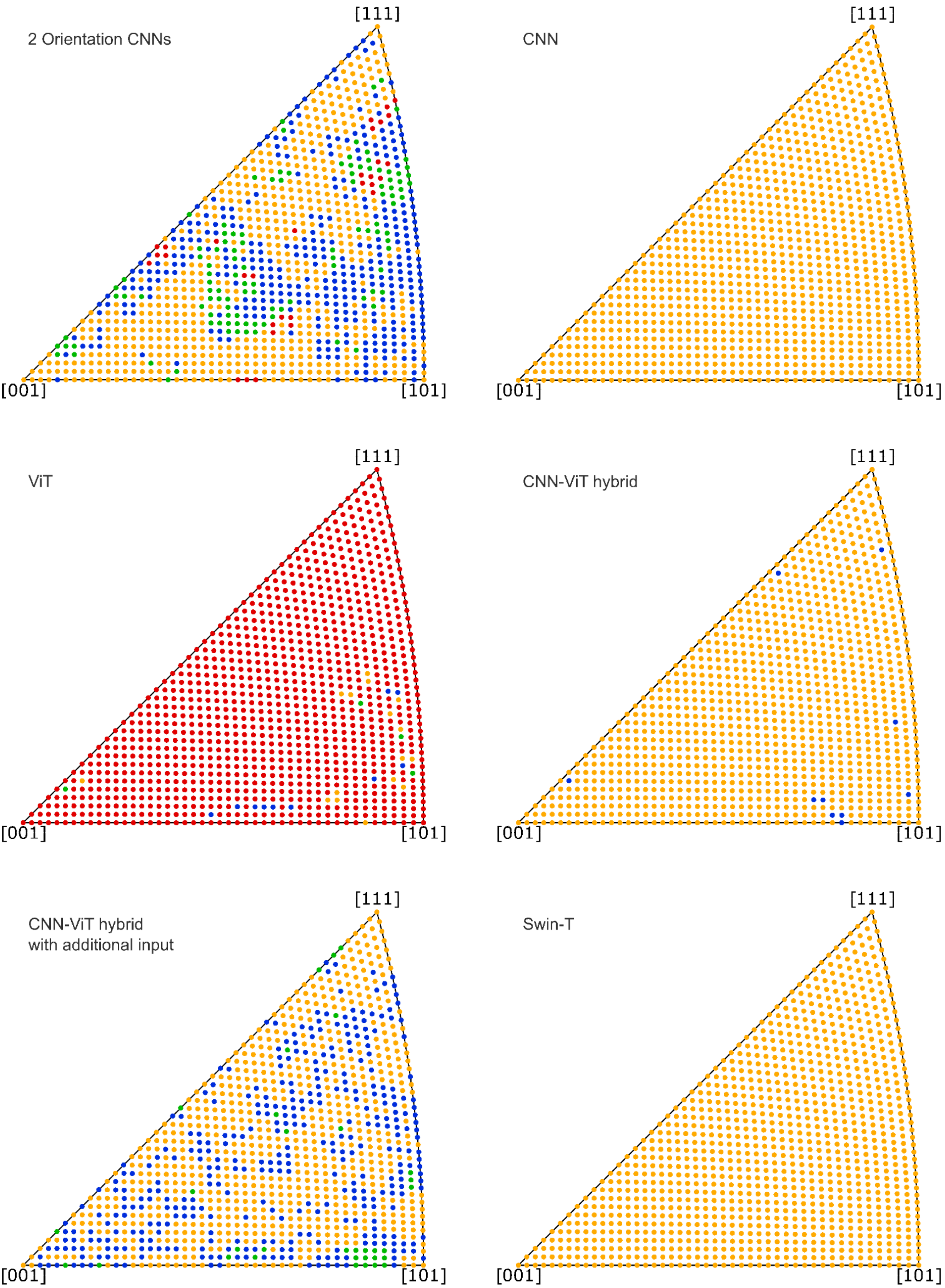

2 Orientation CNNs
[111]
[001]
[101]
CNN
[111]
[001]
[101]
ViT
[111]
[001]
[101]
CNN-ViT hybrid
[111]
[001]
[101]
CNN-ViT hybrid
with additional input
[111]
[001]
[101]
Swin-T
[111]
[001]
[101]

Fig S3: **Phase prediction on NiO data.** Visualisation of predictions made of the networks on non-augmented synthetic NiO images for all orientations in the training dataset (excluding phi1 variation). Blue: LNO, Green: NiO, Yellow: amorphous, Red: vacuum.

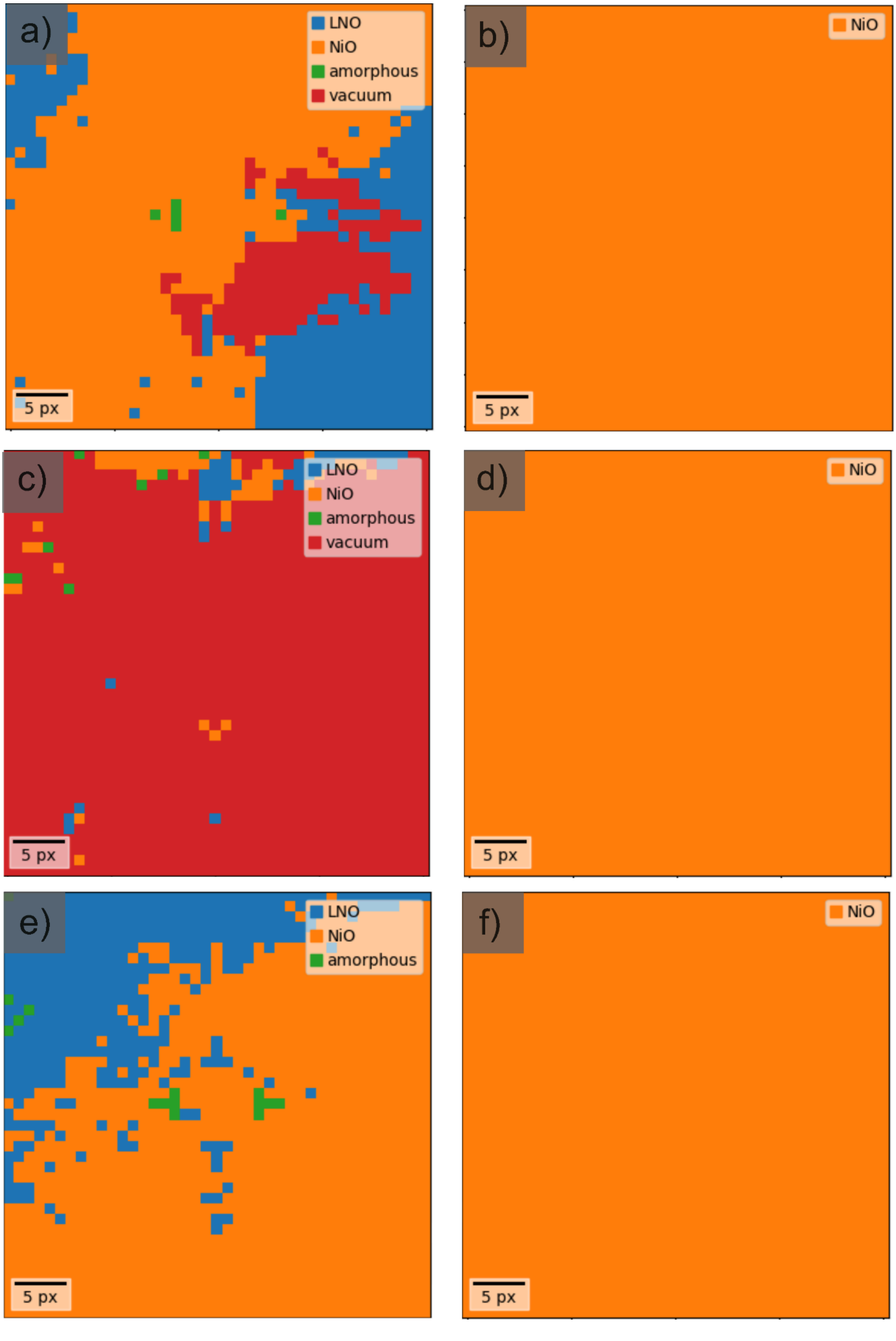

a)
LNO
NiO
amorphous
vacuum
5 px
b)
NiO
5 px
c)
LNO
NiO
amorphous
vacuum
5 px
d)
NiO
5 px
e)
LNO
NiO
amorphous
5 px
f)
NiO
5 px

Fig S4: **Predictions around challenging orientations.** Phase predictions in steps of 0.2 degrees around the orientation [270°, 90°, 90°] of NiO. Movement in the x-Axis represents a shift for PHI while a movement in the y-axes represents a shift for phi2. a) resulting predictions from comparing two separate orientation networks, b) prediction of the CNN, c) prediction of the ViT, d) prediction of the CNN-ViT hybrid, e) prediction of the CNN-ViT hybrid with the original image as additional transformer input, f) prediction of the Swin-transformer.

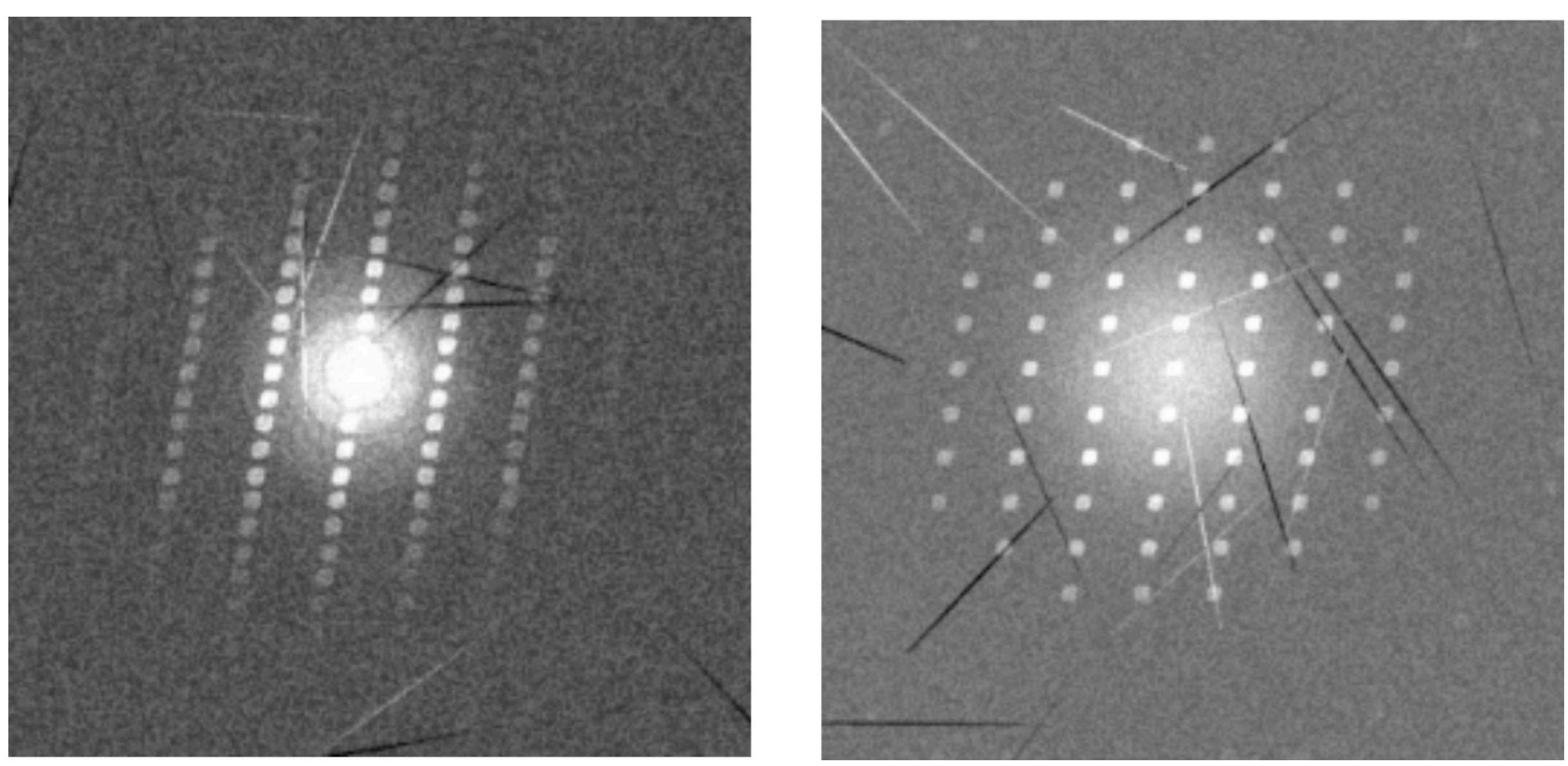

Fig S5: **Highly distorted diffraction patterns.** Examples of LNO test diffraction patterns with heavy augmentations, i.e., shear and Kikuchi lines, which are not present in the training dataset.

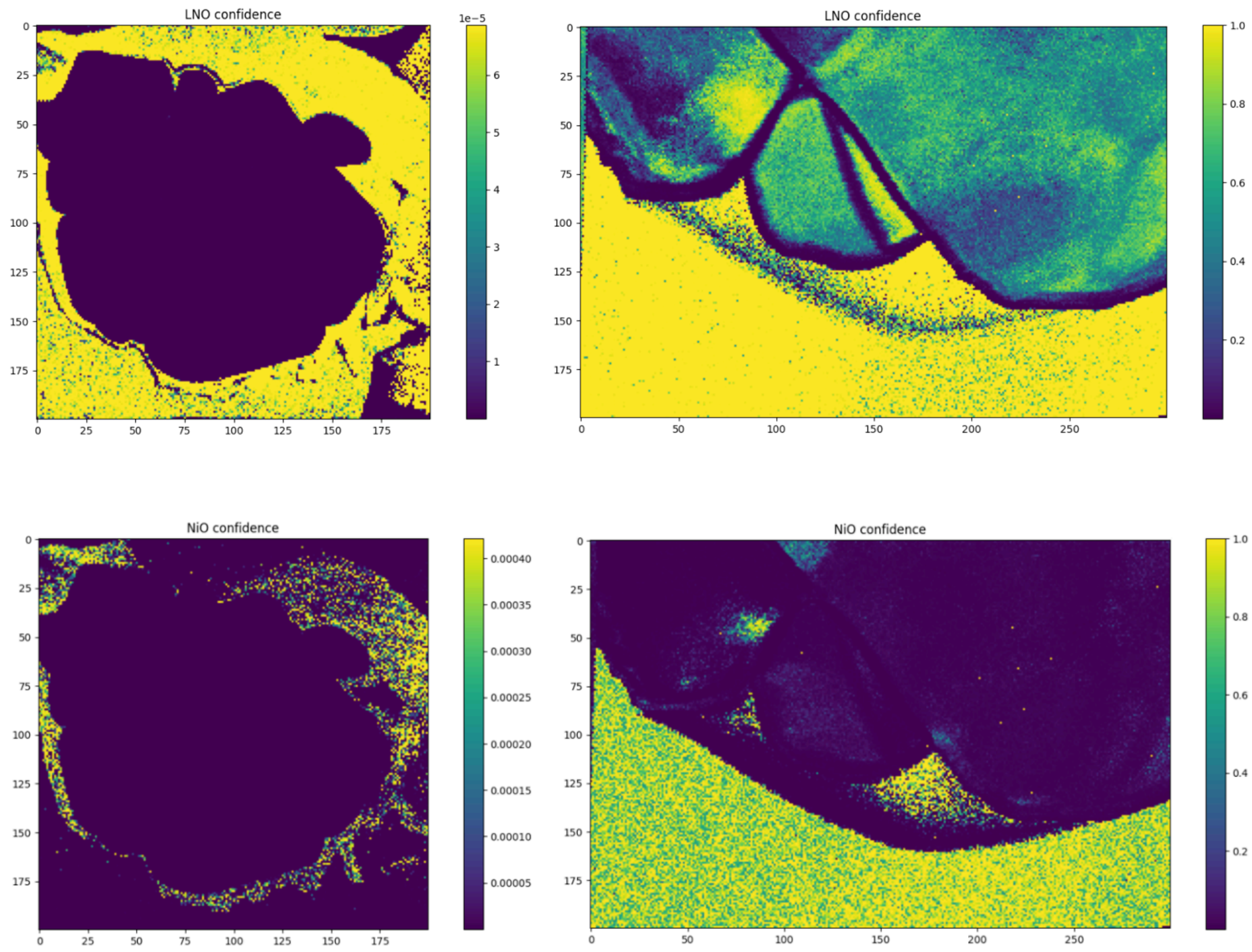

Fig S6: **Orientation network confidences.** Confidence of the separate orientation networks when predicting on the two experimental LNO validation datasets. Left: pure LNO sample. Right: LNO sample with a rock-salt outer edge. Top: Confidence of the LNO network. Bottom: Confidence of the NiO network.

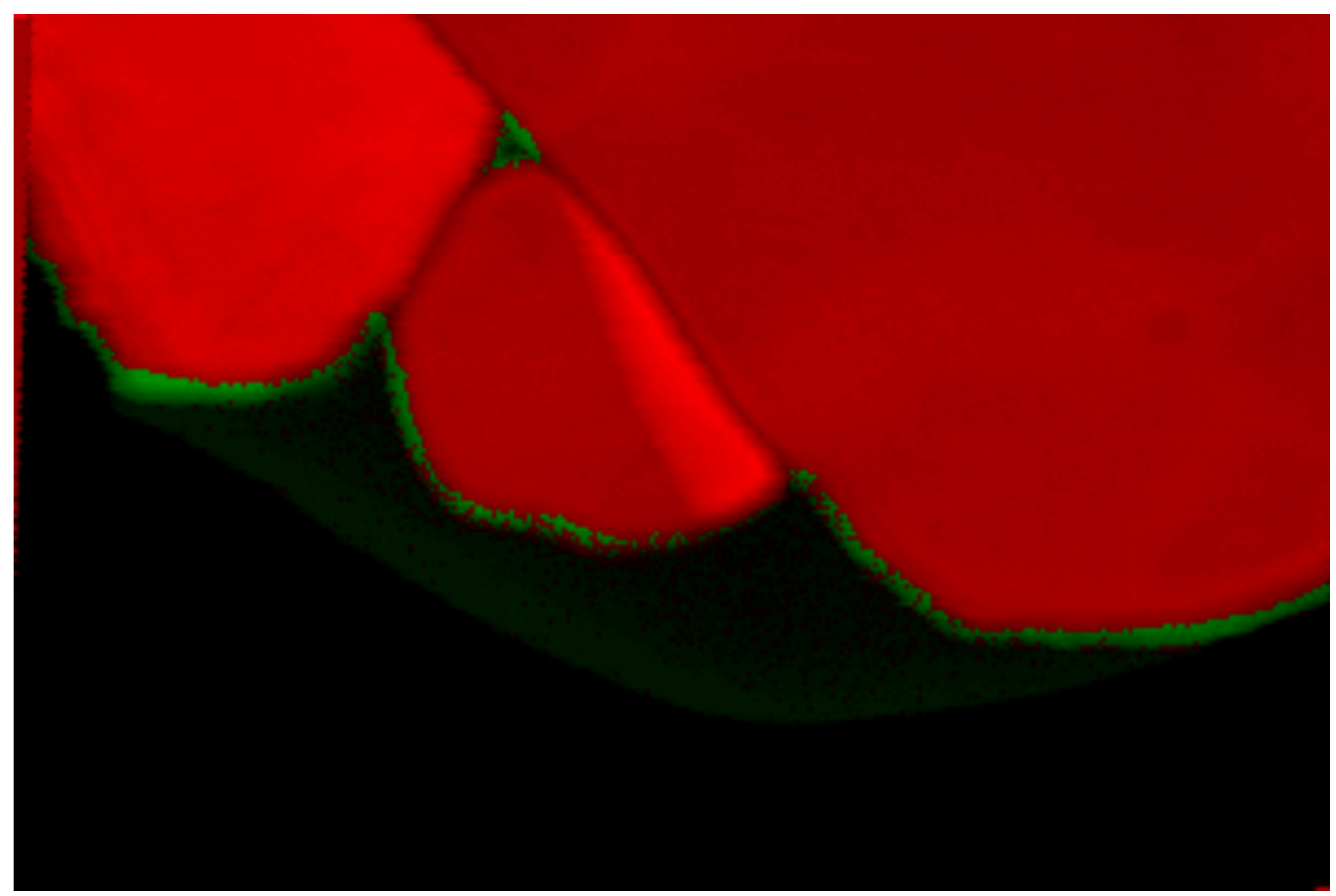

S7: **Reference pattern matching phase mapping.** Phase mapping of the second experimental LNO sample achieved with the ASTAR pattern matching program. Red: LNO phase, Green: NiO phase

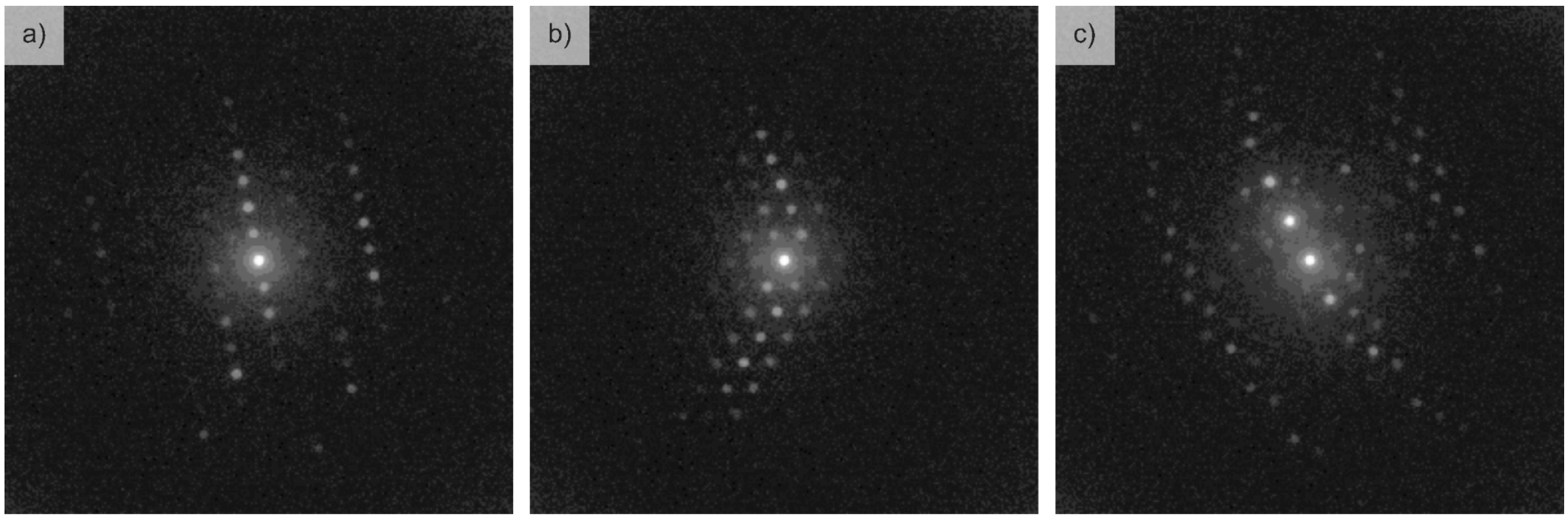

Fig S8: **NiO diffraction patterns.** Exemplary diffraction patterns of the third experimental dataset. As this NiO consists of many overlapping grains, only individually selected diffraction patterns are analysed.

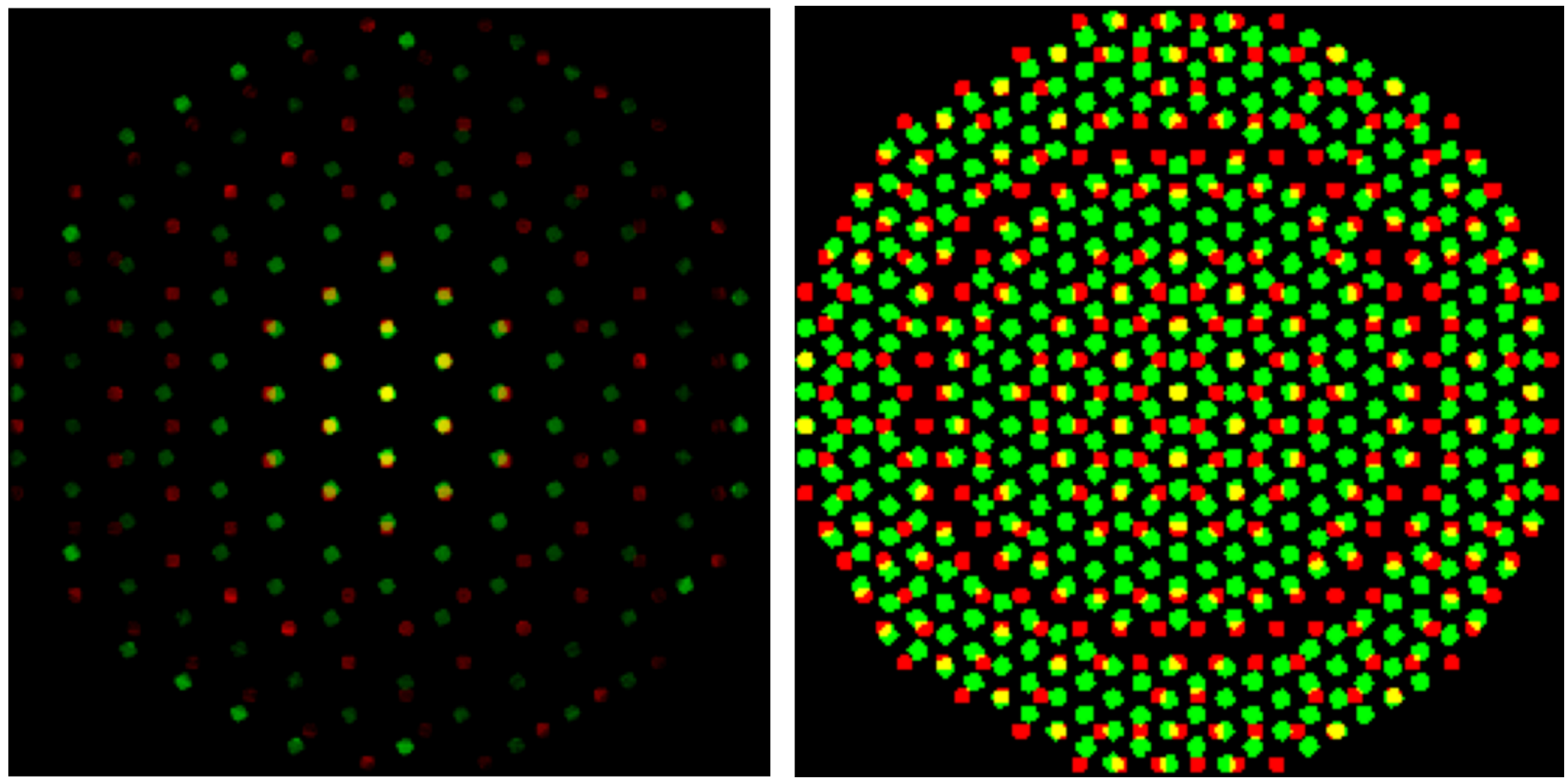

Figure S9: **Excitation error pattern change.** Overlay of the diffraction patterns of LNO in [0001] and NiO in [111] orientation. Cutoff threshold for Ewald-sphere distance reduced for LNO to include a gap between the Laue zones. Red: LiNiO2, Green: NiO. a) diffraction spots scaled with original intensity, b) binary intensity to make every spot clearly visible.

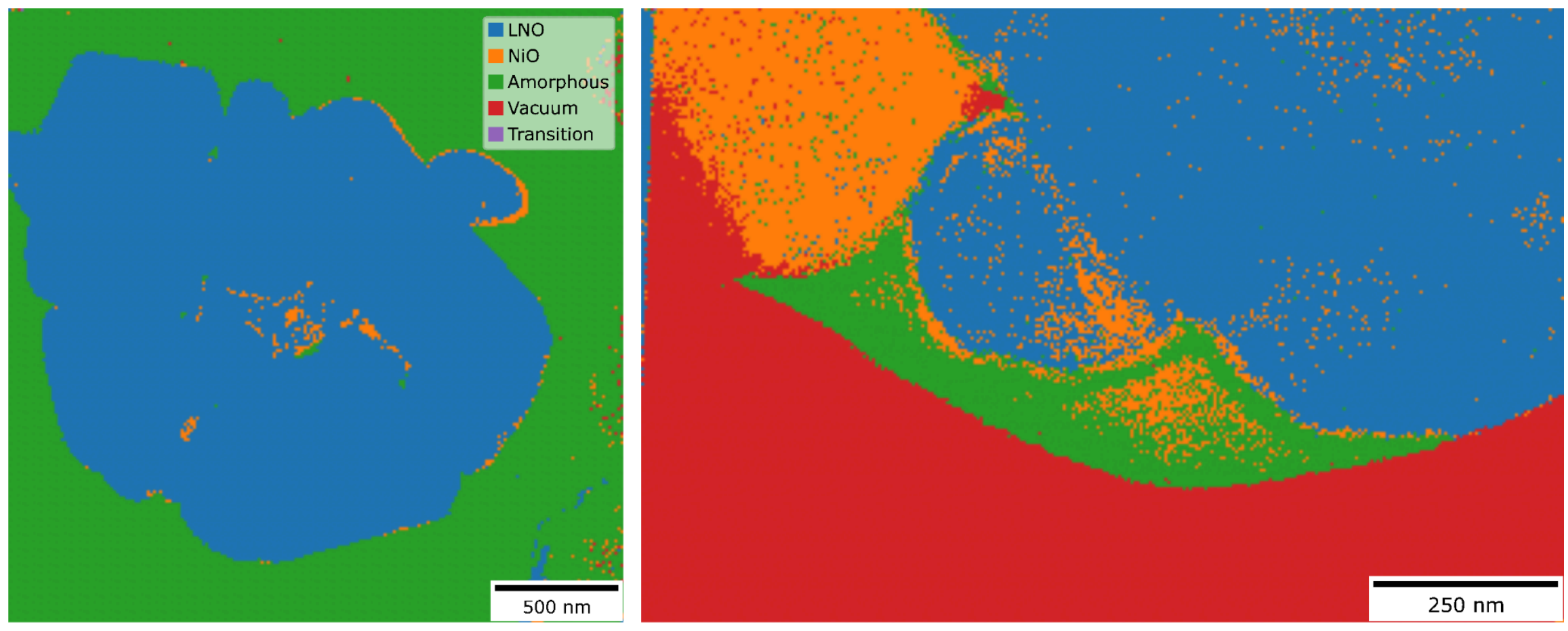

Fig. S10: **Excitation error variation results.** Predictions for the two experimental LNO datasets by a Phase-CNN trained on a synthetic dataset which varies in its excitation error threshold.

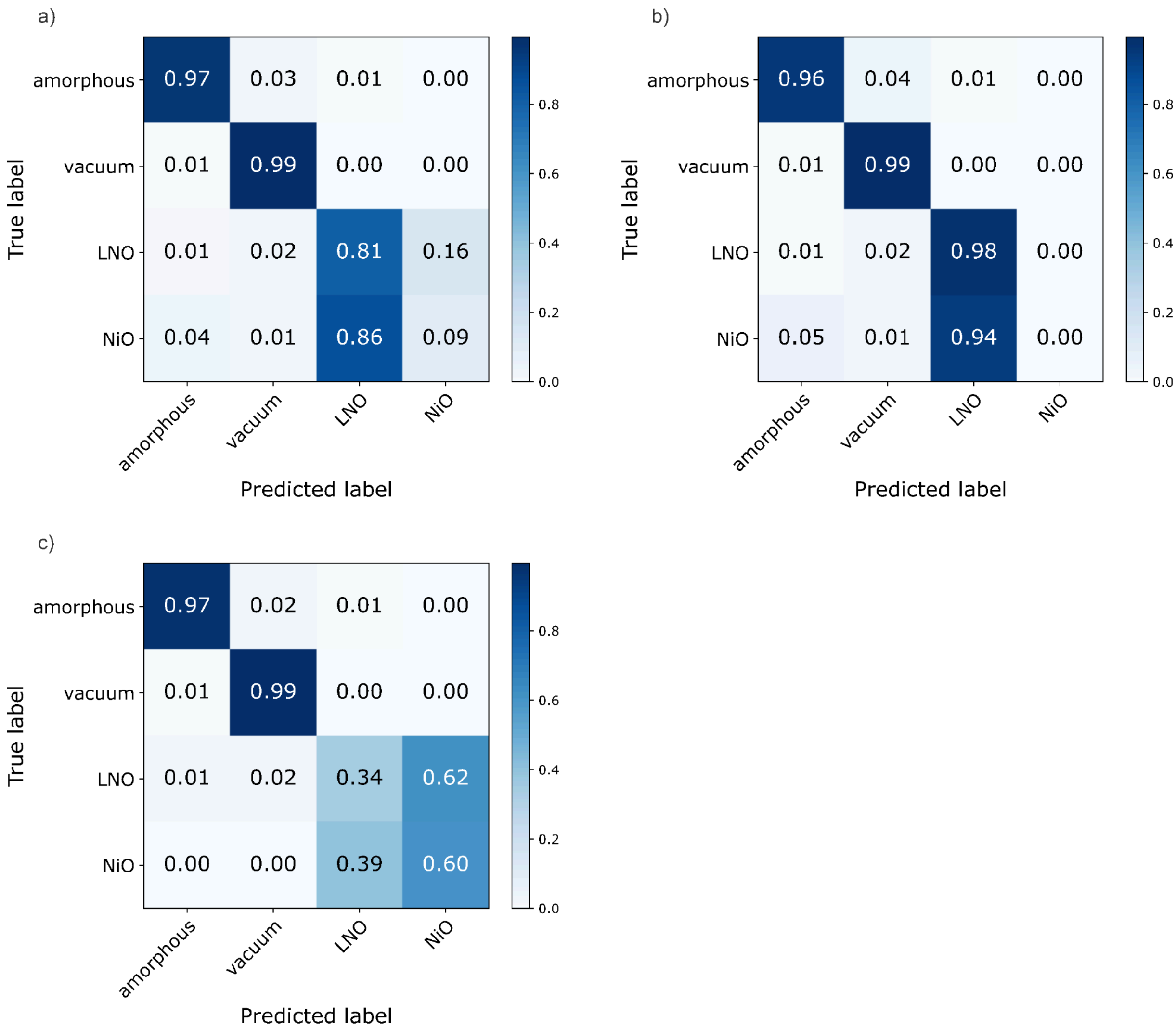

a)

| True label \ Predicted label | amorphous | vacuum | LNO | NiO |
|---|---|---|---|---|
| amorphous | 0.97 | 0.03 | 0.01 | 0.00 |
| vacuum | 0.01 | 0.99 | 0.00 | 0.00 |
| LNO | 0.01 | 0.02 | 0.81 | 0.16 |
| NiO | 0.04 | 0.01 | 0.86 | 0.09 |

b)

| True label \ Predicted label | amorphous | vacuum | LNO | NiO |
|---|---|---|---|---|
| amorphous | 0.96 | 0.04 | 0.01 | 0.00 |
| vacuum | 0.01 | 0.99 | 0.00 | 0.00 |
| LNO | 0.01 | 0.02 | 0.98 | 0.00 |
| NiO | 0.05 | 0.01 | 0.94 | 0.00 |

c)

| True label \ Predicted label | amorphous | vacuum | LNO | NiO |
|---|---|---|---|---|
| amorphous | 0.97 | 0.02 | 0.01 | 0.00 |
| vacuum | 0.01 | 0.99 | 0.00 | 0.00 |
| LNO | 0.01 | 0.02 | 0.34 | 0.62 |
| NiO | 0.00 | 0.00 | 0.39 | 0.60 |

Fig S11: **Other orientation network comparison methods.** Confusion matrices generated by comparing different output methods of the two separate orientation CNNs. a) comparing the energy, b) comparing the entropy, c) comparing the logits directly.